\documentclass[aps,prx,10pt,twocolumn,superscriptaddress,notitlepage,floatfix]{revtex4-2}

\usepackage{lmodern}

\usepackage{graphicx}% Include figure files
\graphicspath{{figs_main/}{figs_SI_theory/}{figs_SI_experiment/}}

\usepackage{dcolumn}% Align table columns on decimal point
\usepackage{bm}% bold math
\usepackage{subfigure}
\usepackage{color}
\usepackage{soul}
\usepackage[dvipsnames]{xcolor}

\usepackage{amsfonts,verbatim}

\usepackage[margin=0.6in]{geometry}

\usepackage{tikz}

\usepackage{times}
\usepackage{dsfont}
\usepackage{latexsym}
\usepackage{float}
\usepackage{afterpage}
\usepackage{enumitem}
\usepackage{mleftright}
\usepackage[normalem]{ulem}

\definecolor{lR}{rgb}{1, 0.8, 0.79}

\usepackage{eso-pic, graphicx}

\usepackage{listings}
\usepackage{multirow}
\usepackage{xcolor,colortbl}
\usepackage[dvipsnames]{xcolor}

\usepackage{bbm}
\usepackage{upgreek}
\usepackage{newtxtext,newtxmath}

\newcommand{\nocontentsline}[3]{}
\newcommand{\tocless}[3]{
\vspace{1em}
\bgroup\let\addcontentsline=\nocontentsline#1{#2\label{#3}}\egroup
}

\definecolor{Ablue}{rgb}{0.96,0.24,0.00}

\definecolor{Abluetitle}{rgb}{0.,0.24,0.51}

\definecolor{orange}{rgb}{0.96,0.24,0.00}

\definecolor{darkred}{rgb}{0.55, 0.0, 0.0}

\definecolor{darksalmon}{rgb}{0.91, 0.59, 0.48}
\definecolor{maroon}{cmyk}{0,0.87,0.68,0.32}

\setcitestyle{numbers,square,citesep={,\kern-.24em}}

\definecolor{mustard}{rgb}{1.0, 0.86, 0.35}

\definecolor{Gray}{gray}{0.85}
\definecolor{LightCyan}{rgb}{0.88,1,1}
\newcolumntype{a}{$>${\columncolor{Gray}}c}
\newcolumntype{b}{$>${\columncolor{White}}c}

\usepackage{array}
\newcolumntype{L}[1]{$>${\raggedright\let\newline\\\arraybackslash\hspace{0pt}}m{#1}}
\newcolumntype{C}[1]{$>${\centering\let\newline\\\arraybackslash\hspace{0pt}}m{#1}}
\newcolumntype{R}[1]{$>${\raggedleft\let\newline\\\arraybackslash\hspace{0pt}}m{#1}}

\newcolumntype{P}[1]{>{\centering\arraybackslash}p{#1}}
\newcolumntype{M}[1]{>{\centering\arraybackslash}m{#1}}

\usepackage[
    colorlinks=true,
    citecolor=black,
    urlcolor=black,
    linkcolor=black,
    filecolor=black
]{hyperref}
\newcommand{\id}{\mathds{1}}

\newcommand{\tm}{{\text -}}

\newcommand{\tacq}{t_{\R{acq}}}

\newcommand{\xt}{\vartheta}

\newcommand{\xr}{\rho}
\newcommand{\xo}{\omega}
\newcommand{\xph}{\phi}

\newcommand{\app}{\approx}

\newcommand{\Bpol}{\textbf{B}_{\R{pol}}}

\newcommand{\Bp}{B_{\R{pol}}}

\newcommand{\Cs}{{}^{13}\R{C}}

\newcommand{\Bsens}{B_{\R{sens}}}

\newcommand{\fac}{f_{\R{AC}}}

\newcommand{\Bac}{B_{\R{AC}}}

\newcommand{\mHdd}[0]{\mH_{\R{dd}}}

\newcommand{\mHeff}[0]{\mH_{\R{eff}}}

\newcommand{\xy}[0]{\xhat\tm\yhat}

\newcommand{\xD}{\Delta}

\newcommand{\mH}[0]{\mathcal{H}}

\newcommand{\beq}{\begin{equation}}
\newcommand{\eeq}{\end{equation}}
                  
\newcommand{\benum}{\begin{enumerate}}
\newcommand{\eenum}{\end{enumerate}}
                    
\newcommand{\bit}{\begin{itemize}}
\newcommand{\eit}{\end{itemize}}
\newcommand{\xhat}{\hat{\T{x}}}
\newcommand{\yhat}{\hat{\T{y}}}
\newcommand{\zhat}{\hat{\T{z}}}
\newcommand{\nhat}{\hat{\T{n}}}

\newcommand{\bea}{\begin{eqnarray}}
\newcommand{\eea}{\end{eqnarray}}

\newcommand{\bev}{\begin{verbatim}}
\newcommand{\eev}{\end{verbatim}}

\newcommand{\zt}{\times}

\newcommand{\qt}{\tau}

\newcommand{\T}[1]{\textbf{#1}}
\newcommand{\I}[1]{\textit{#1}}
\newcommand{\R}[1]{\textrm{#1}}
\newcommand{\zfl}[1]{\protect\label{fig:#1}}
\newcommand{\zfr}[1]{\figurename\,\ref{fig:#1}}

\newcommand{\zsl}[1]{\label{sec:#1}}
\newcommand{\zsr}[1]{Sec.\,\ref{sec:#1}}
\newcommand{\ba}{\left\{ \begin{array}{lr}}
\newcommand{\ea}{\end{array}\right.}

\newcommand{\blist}[1]{
 \begin{list}{#1}%$\ast\circ\bullet\Right
 \begin{align}
	 arrow
 \end{align}
 $\checkmark\star
  { \setlength{\itemsep}{3pt}
     \setlength{\parsep}{2pt}
     \setlength{\topsep}{3pt}
     \setlength{\partopsep}{0pt}
     \setlength{\leftmargin}{1em}
     \setlength{\labelwidth}{1em}
     \setlength{\labelsep}{0.5em} } }
\newcommand{\elist}{
  \end{list}  }

\DeclareMathSymbol{\vartheta}{\mathalpha}{letters}{"12}
\DeclareMathSymbol{\theta}{\mathalpha}{letters}{"23}
\DeclareMathSymbol{\phi}{\mathalpha}{letters}{"27}
\DeclareMathSymbol{\varphi}{\mathalpha}{letters}{"1E}

\newcommand{\bef}
{
\begin{figure}[htbp]
\centering
}

\newcommand{\eef}{\end{figure}}

\mleftright
\makeatletter
\@addtoreset{section}{part}
\makeatother

\newcommand{\beginsupplement}{%
        \setcounter{table}{0}
        \renewcommand{\tablename}{Supplementary Table}
        \renewcommand{\thetable}{\arabic{table}}%
        \setcounter{figure}{0}
        \renewcommand{\thefigure}{S\arabic{figure}} %
        \renewcommand{\theHfigure}{S\arabic{figure}} %fixes linking to figures
		\setcounter{page}{1}
		\renewcommand{\figurename}{Fig.} 
		\renewcommand{\thesection}{\:S\arabic{section}}
		\setcounter{section}{0}
        \setcounter{equation}{0}
        \renewcommand{\theequation}{S\,\arabic{equation}}
     }

\newcommand{\affA}{Department of Chemistry, University of California, Berkeley, Berkeley, CA 94720, USA.}

\newcommand{\affD}{John A. Paulson School of Engineering and Applied Sciences, Harvard University, Cambridge, MA 02138, USA.}
\newcommand{\affE}{Chemical Sciences Division,  Lawrence Berkeley National Laboratory,  Berkeley, CA 94720, USA.}
\newcommand{\affF}{CIFAR Azrieli Global Scholars Program, 661 University Ave, Toronto, ON M5G 1M1, Canada.}

\begin{document}
\title{Anomalously extended Floquet prethermal lifetimes and applications to long-time quantum sensing}
\author{Kieren A Harkins}\thanks{These authors contributed equally to this work}\affiliation{\affA}
\author{Cooper Selco}\thanks{These authors contributed equally to this work}\affiliation{\affA}\affiliation{\affE}
\author{Christian Bengs}\affiliation{\affA}\affiliation{\affE}
\author{David Marchiori}\affiliation{\affA}
\author{Leo Joon Il Moon}\affiliation{\affA}\affiliation{\affE}
\author{Zhuo-Rui Zhang}\affiliation{\affA}
\author{Aristotle Yang}\affiliation{\affA}
\author{Angad Singh}\affiliation{\affA}
\author{Emanuel Druga}\affiliation{\affA}
\author{Yi-Qiao Song}\affiliation
{\affD}
\author{Ashok Ajoy}\email{ashokaj@berkeley.edu}\affiliation{\affA}\affiliation{\affE}\affiliation{\affF}

\begin{abstract}
Floquet prethermalization is observed in periodically driven quantum many-body systems where the system avoids heating and maintains a stable, non-equilibrium state, for extended periods. Here we introduce a novel quantum control method using off-resonance and short-angle excitation to significantly extend Floquet prethermal lifetimes. This is demonstrated on randomly positioned, dipolar-coupled, $\Cs$ nuclear spins in diamond, but the methodology is broadly applicable. We achieve a lifetime $T_2' {\app
} 800$ s at 100 K while tracking the transition to the prethermal state quasi-continuously. This corresponds to a ${>}$533,000-fold extension over the bare spin lifetime without prethermalization, and constitutes a new record both in terms of absolute lifetime as well as the total number of Floquet pulses applied (here exceeding 7 million). Using Laplace inversion, we develop a new form of noise spectroscopy that provides insights into the origin of the lifetime extension. Finally, we demonstrate applications of these extended lifetimes in long-time, reinitialization-free quantum sensing of time-varying magnetic fields continuously for ${\sim}$10 minutes at room temperature. Our work facilitates new opportunities for stabilizing driven quantum systems through Floquet control, and opens novel applications for continuously interrogated, long-time responsive quantum sensors.
 
\end{abstract}

\maketitle
\pagebreak
\begin{figure}[t]
  \centering
 {\includegraphics[width=0.49\textwidth]{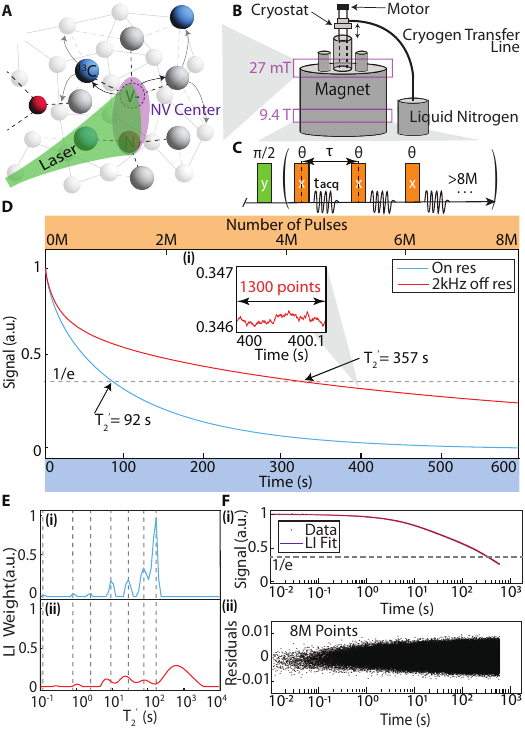}}
    \caption{\T{System and principle.} 
(A) \I{System:} diamond NV electrons coupled to $\Cs$ nuclei.
(B) \I{Apparatus:} Hyperpolarization is carried out at low field ($\Bp{\app}$27mT), and $\Cs$ control and readout is performed at high-field $B_0$. Sample is contained in a cryostat shuttled between the two field centers.
(C) \I{$\Cs$ Interrogation} at $B_0{=}9.4$T consists of a train of spin-locking $\xt$ pulses (length $t_p$), with $^{13}$C Larmor precession interrogated in $\tacq$ windows between them. 
(D) \I{Extension of $T_2’$ with resonance offset} at 100K. Blue and red lines show normalized data measured on- and 2kHz off-resonance respectively with $T_2’{=}92$s and $T_2’{=}357$s (${\sim}$6min) respectively obtained from a $1/e$ intercept. Dashed line shows the $1/e$ value. Each trace has ${>}8$M points (upper axis).
\I{Inset (i)}: Zoom into a $\xD t{=}100$ms window at $t{=}400$s, with 1300 points.
(E) \I{Laplace inversion (LI)} carried out for data in (D) unravelling $T_2'^{j}$ components over five orders of magntiude in time. Dashed lines show peak components corresponding to Fig. 1E(i). For the off-resonance case, long-time component is significantly extended to $T_2’^{j}{\app}680$s. 
(F) \I{Efficacy of LI fit.} (i) Red points show data corresponding to Fig. 1E(i) on semi-log scale, while purple is the LI derived fit. (ii) \I{Bottom panel:} fit residuals, showing a ${<}1$\% variation over the entire 600s period and ${>}$8M points.}
\zfl{mfig1}
\end{figure}

\begin{figure*}[t]
  \centering
  {\includegraphics[width=\textwidth]{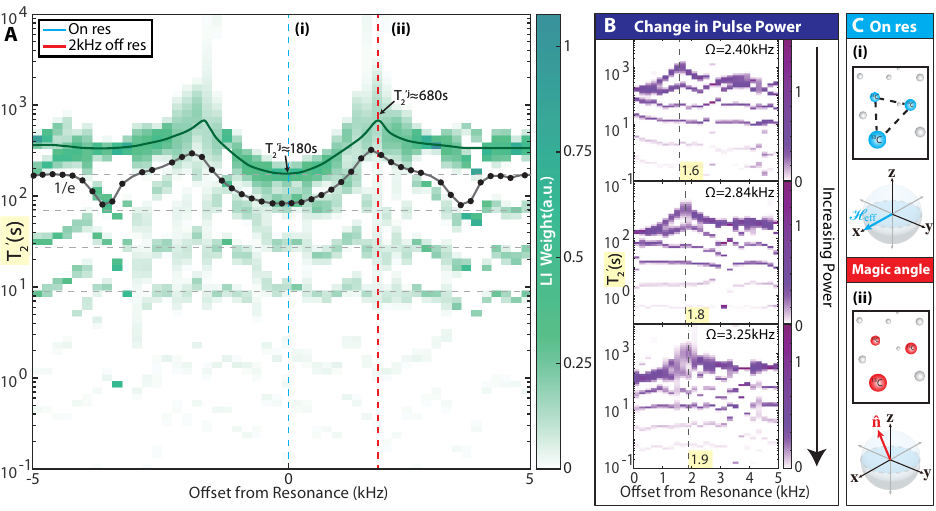}}
     \caption{\T{$T_2’$ lifetime extension due to frequency offset.} 
(A) \I{Laplace Inversion (LI) map} constructed from a dataset with 51 values of $\Delta\xo\in [-5,5]$ kHz. Here $t_p{=}34\mu$s, $\qt{=}43\mu$s, and $\tacq{=}4\mu$s in \zfr{mfig1}C. 
%Extracted $T_2'^{j}$ values are plotted on the vertical axis
Colors represent the strength of constituent LI weights $w_j$ (color bar).
Extension of long-time $T_2'^{j}$ component at $\xD\xo{=}\pm2$ kHz is evident, yielding two horn-like features with $T_2'^{j}{\app}680$ s. Shorter time components remain consistent across offsets (gray dashed lines). Black points indicate $T_2'$ extracted from a $1/e$ intercept; solid green line is a guide to eye.
\I{Dashed vertical lines} show representative line cuts at \I{(i)} $\Delta\xo{=}0$ and \I{(ii)} $\Delta\xo{=}2$ kHz, corresponding to \zfr{mfig1}D.
(B) \I{Effect of changing Rabi frequency $\Omega$}: LI maps with increasing pulse power and fixed nominal $\xt{=}90^{\circ}$. Horns move to larger offset values $\Delta\xo$ (dashed vertical lines and highlighted numbers) , consistent with Lee-Goldberg decoupling.
(C) \I{Schematic.}  \I{(i)} On-resonance, the $\Cs$ nuclei prethermalize to effective Hamiltonian $\mHeff$ aligned with the RF drive along $\xhat$. Dashed lines indicate dipolar couplings. \I{(ii)} Off-resonance at the LG condition, $\Cs$ nuclei prethermalize along magic angle $\nhat$ and are self-decoupled to zeroth order.}
\zfl{mfig2}
\end{figure*}

\begin{figure*}[t]
  \centering
  {\includegraphics[width=\textwidth]{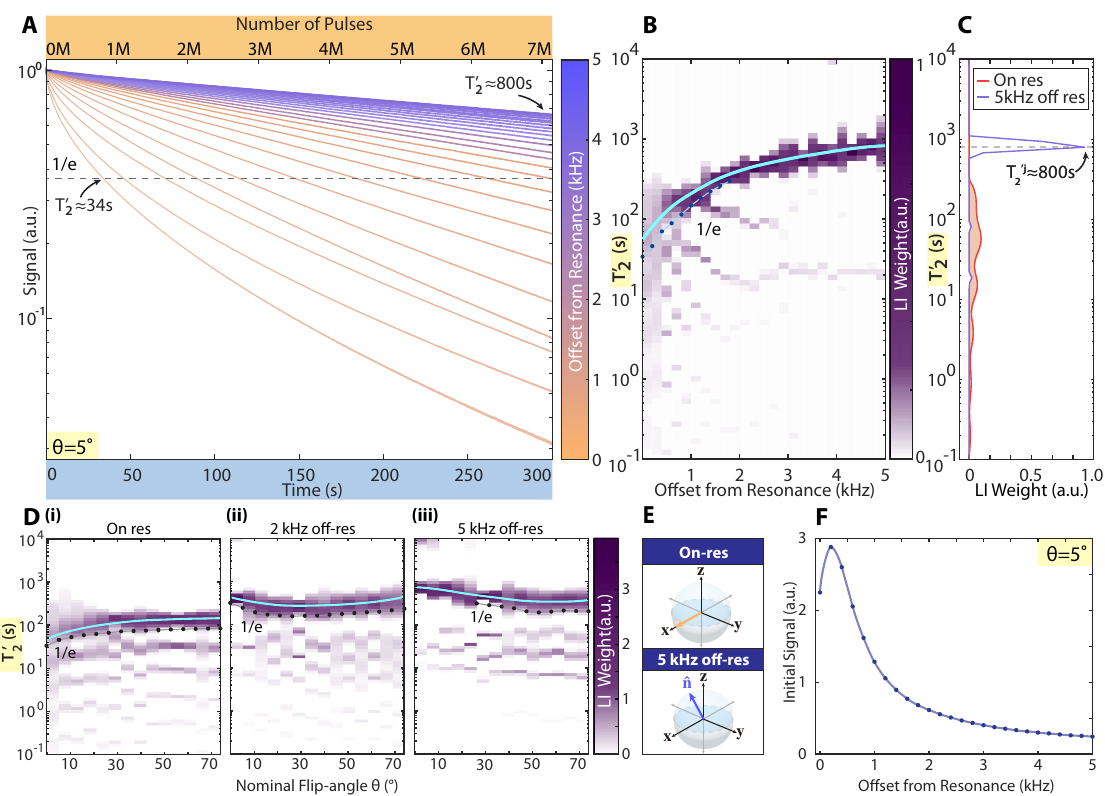}}
       \caption{
\T{Anomalously long $T_2'$ lifetimes with small-angle pulses.}
(A) \I{Normalized decay profiles} on a semi-log scale for 26 $\xD\xo$ offset values (see colorbar) with nominal $\xt{=}5^\circ$. Each trace has ${>}7$M pulses (upper axis). Dashed line shows $1/e$ intercept; we estimate $T_2'{\app}800$s for $\Delta\xo{=}5$ kHz. 
(B) \I{Laplace Inversion (LI) map} for data in (A) demonstrates that curves become increasingly mono-exponential with increasing offset. Solid cyan line is a guide to eye for the longest $T_2^{'j}$ component. Black points are corresponding $T_2'$ values extracted from a $1/e$ intercept.
(C) \I{Representative LI traces} for curves (i) on-resonance (orange shaded) and (ii) $\Delta\xo{=}5$ kHz off-resonance (purple). The latter shows a single component at $T_2^{'j}{\app}800$s.
(D) \I{LI maps} of normalized traces for varying (nominal) $\xt$ under three different offset conditions. Solid cyan lines are guide to eye on the long-time component. Black points are derived from $1/e$ intercepts.
\I{(i)} On resonance, $T_2'$ values decrease as expected with lowering $\xt$.
\I{(ii)} At intermediate offsets, $\xD\xo{=}2$ kHz, $T_2'$ values increase moderately at short $\xt$.
\I{(iii)} At large offsets, $\xD\xo{=}5$ kHz, $T_2'$ values rise anomalously at short $\xt$.
(E) \I{Schematic Bloch-sphere description} on- and off-resonance at short $\xt$.
(F) \I{Amplitude change} for the initial signal $S$ as a function of offset for $\xt{=}5^\circ$. Solid blue line is a guide to eye. Extension in $T_2'$ lifetime far exceeds the loss in signal.}
\zfl{mfig3}
\end{figure*}

\begin{figure*}[t]
  \centering
  {\includegraphics[width=\textwidth]{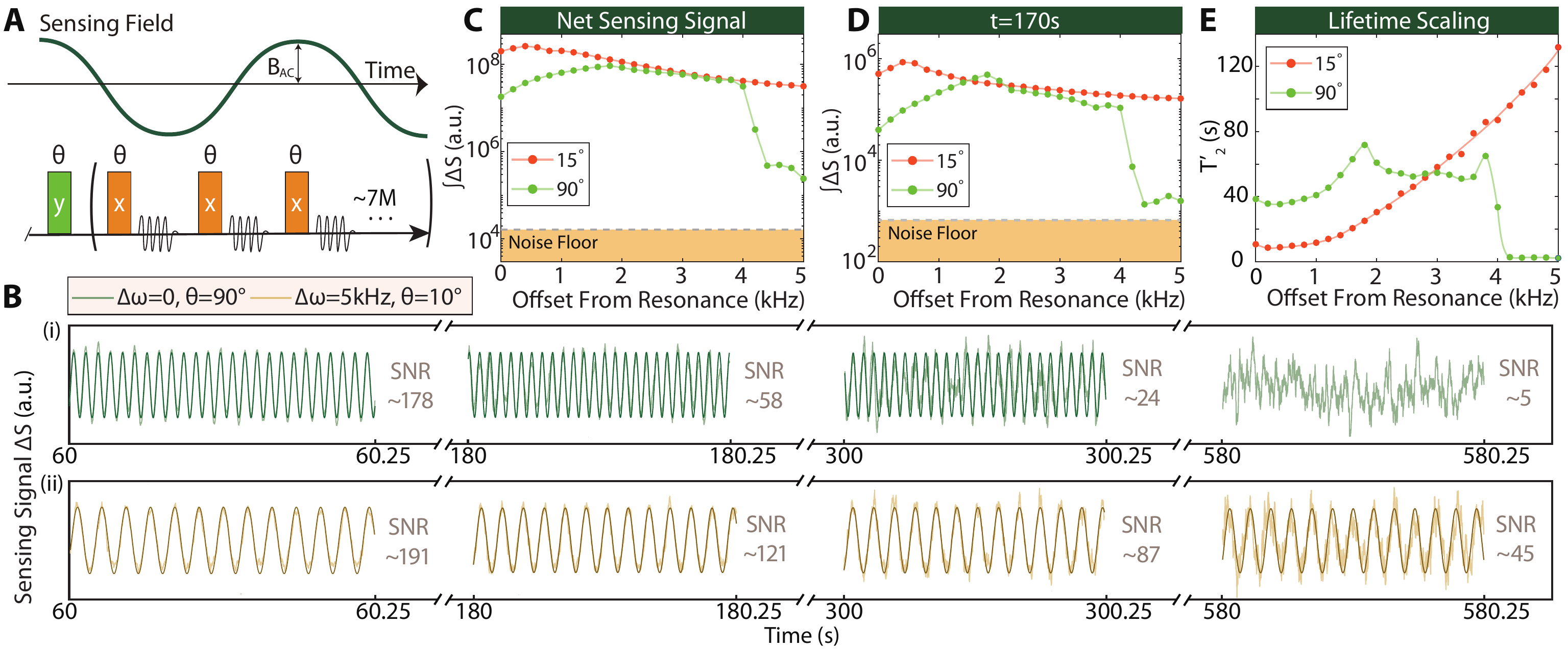}}
 \caption{\T{Long-time quantum sensing at room temperature.}
(A) \I{Sensing protocol} with a pulse sequence similar to \zfr{mfig1}C. Spins are simultaneously exposed to sensing AC field with frequency $f_{AC}$ and amplitude $B_{AC}$.
(B) \I{Comparison of sensing signal $\Delta S$} for $\Bac{=}82\mu$T and $\fac{=}50$Hz (i) on-resonance and $\xt{=}90^\circ$ (green), and (ii) $\xD\xo{=}5$kHz off-resonance and nominal $\xt{=}10^\circ$ (yellow). Light line is data, dark line is a sinusoidal fit. Shown are four $\xD t{=}100$ms windows centered at $60$s, $180$s, $300$s, and $580$s; marked in gray is SNR in each window calculated over 1s. Data is smoothed using a boxcar average of 30 points. Off-resonance small-angle sensing signal starts at a higher level and persists significantly longer. 
(C) \I{Integrated signal $\int\Delta S(t')dt'$} as a function of offset $\Delta\omega$ for nominal flip angles $\xt{=}15^{\circ}$ and $\xt{=}90^{\circ}$ for data measured over $t{=}$180s. This is measured from sum of the absolute value of the $\xD S$ signal Fourier transform at $\fac$ and $2\fac$. Shaded region denotes Johnson noise floor.
(D) \I{Integrated signal} for a $\xD t{=}1$s window at $t{=}$170s.
(E) \I{Lifetime $T'_2$} as a function of offset $\Delta\omega$. Lifetime is significantly extended far off-resonance for $\xt{=}15^{\circ}$.
}
\zfl{mfig4}
\end{figure*}

\vspace{-1cm}
\section{Introduction}
Floquet prethermalization -- the arrest of thermalization in periodically driven many-body quantum systems -- has garnered significant recent attention ~\cite{abanin2015exponentially,kuwahara2016floquet,weidinger2017floquet,else2017prethermal}.  Periodic kicks induce the system to enter a long-lived metastable state rather than heating rapidly, enabling applications that exploit many-body couplings before the system ultimately thermalizes to infinite temperature~\cite{singh2019quantifying, beatrez2021floquet, peng2021floquet,rubio2020floquet}. Among these are the formation of non-equilibrium states like prethermal time crystals \cite{choi2017observation,zhang2017observation,  rovny2018observation, kyprianidis2021observation, beatrez2023critical, moon2024experimental}, stabilized topological phases \cite{potirniche2017floquet, ye2021floquet, zhang2022digital}, and the creation of stabilized nanoscale spin textures \cite{harkins2023nanoscale}. It has also been exploited for quantum sensing \cite{sahin2022high}, wherein the prethermal state is rendered responsive to external fields; the long-lifetime boosting sensitivity and resolution.

A key focus for these applications, especially in sensing, is on developing strategies to extend prethermal lifetimes via Floquet control~\cite{dong2008controlling, fleckenstein2021thermalization, zhao2022suppression}, and understanding the origins of heating processes during different stages of thermalization \cite{ho2023quantum,weidinger2017floquet, mallayya2019prethermalization, beatrez2021floquet,birnkammer2022prethermalization}.

Consider nuclear spins in a solid coupled via magnetic dipolar interactions. When prepared in a superposition state, $\xr_0{\propto}I_x$, where $I$ is the net spin-1/2 Pauli operator, the state rapidly decays to infinite temperature $\xr_{\infty}{\sim}\id$ with a characteristic time-scale $T_2^*$, driven by the interactions. However, suitable Floquet control can reengineer the dipolar Hamiltonian to $\mHeff$, where $[\mHeff,I_x]{=}0$, inducing prethermalization, and significantly extending the state's survivial. For $\Cs$ nuclei in diamond, we demonstrated a lifetime extension from \( T_2^* {\approx} 1.5 \) ms to \( T_2' {=} 90 \) s \cite{beatrez2021floquet}. Lifetimes are typically estimated from a $1/e$-intercept of the decay profiles, but this approach, while useful, is simplistic and incomplete.

Several key questions arise: (i) can these lifetimes be further extended, (ii) can broader insights into the thermalization process be gained by examining the heating profiles to infinite temperature, and (iii) can this be leveraged for sensing? In this study, using optically hyperpolarized $^{13}$C nuclei and new Floquet control strategies, we address these aspects. 

Notably, we observe that prethermal lifetimes can be significantly increased through tailored off-resonant excitation combined with short-angle pulses (nominally $\xt{<}10^\circ$), extending the $1/e$-lifetime to \mbox{\( T_2' {>}800 \) s} -- a $>$533,000-fold increase over $T_2^*$ under identical conditions. This corresponds to over 7M control pulses applied to the spins, setting a new record for both the duration of prethermalization and the number of Floquet pulses sustained (see detailed table of comparison in \zfr{SItable}). Remarkably, however, the extension itself is achieved with only a small number of pulses per $T_2^*$ period, leading us to term this an \I{anomalously} extended prethermal lifetime.

By leveraging the high signal-to-noise ratio (SNR) and rapid data sampling (millions of data points) enabled by hyperpolarization and instrumentation advances, we develop a powerful approach based on Laplace inversion (LI)~\cite{song2002t1, venkataramanan2002solving} to quantify individual decay channels in the thermalization process. This provides more detailed insights than simple $1/e$ intercepts and enables a deeper understanding of lifetime extensions via Floquet control.

Building on this, we demonstrate practical applications in long-time, continuously interrogated AC magnetometry with hyperpolarized $^{13}$C nuclei~\cite{sahin2022high}. Magnetic fields are measured continuously for ${\app}$10min post a single initialization step, marking the longest such demonstration of quantum sensing to our knowledge.

\section{Results}
\subsection{System and Principle: Laplace inversion of prethermal dynamics}
\zsl{principle}
\zfr{mfig1}A depicts the system. It consists of a central nitrogen vacancy (NV) electron surrounded by $\Cs$ nuclei at natural abundance. NVs are spaced ${\approx}$24 nm, and $\Cs$ spins are distributed at ${\approx}$0.92/nm$^3$~\cite{van1997dependences}, though in a random manner. They interact via magnetic dipole interactions~\cite{abragam1961principles}, $\mHdd = \sum_{k<l}b_{kl} (3I_{kz} I_{lz} - \T{I}_k\cdot \T{I}_{l})$, with a median strength $J{\app}0.6\,$kHz, where $I_{k\alpha}$ represents the spin-$1/2$ Pauli matrix for the k$^\R{{th}}$ spin, $\alpha{=}\{x,y,z\}$, and net magnetization is $I_{\alpha}=\sum_kI_{k\alpha}$.

Hyperpolarization uses laser and microwave excitation at low magnetic fields ($\Bpol{\app}27$ mT) \cite{ajoy2018orientation, ajoy2021low, sarkar2022rapidly} (\zfr{mfig1}B) and can be performed at temperatures ranging from 100K to room temperature (see Methods). The nuclear spins are subject to Floquet control at high fields, $B_0{\sim}9.4$T, (\zfr{mfig1}B) as outlined in \zfr{mfig1}C, with a series of spin-locking pulses applied after spins are tipped along $\xhat$. Pulse spacing is $\qt$; 
while $\xt$ denotes the flip angle, \I{nominally} assuming pulses are resonant. Spin precession is monitored non-desctuctively during $\tacq$ periods between pulses via an RF cavity(see Methods); the measured signal $S$ is the projection of magnetization onto the $\xy$ plane in the rotating frame (see Methods). 

The blue trace in \zfr{mfig1}D shows the (normalized) signal $S$ with pulses applied on-resonance at 100K. The time-averaged Hamiltonian, $\mHeff=\sum_{j<k}d_{jk}(\frac{3}{2}(I_{jz}I_{kz}+I_{jy}I_{ky})-\Vec{I_{j}}{\cdot}\Vec{I_{k}})$ commutes with the initial state, $[\mHeff,\xr_0]{=}0$ to zeroth order of a Magnus expansion~\cite{mananga2011introduction,kuwahara2016floquet}, leading to prethermalization~\cite{beatrez2021floquet}. The resulting $1/e$ lifetime of $T_2’{\app}92$ s is significantly longer than the free-induction decay time $T_2^{\ast}{=} 1.5$ ms which is dominated decay under the inter-spin dipolar interactions. The upper axis in \zfr{mfig1}D shows the number of Floquet pulses applied (here ${>}$8M). 

The red trace in \zfr{mfig1}D instead shows the measured signal when pulses applied slightly off-resonance, with offset $\xD\xo{=}$2 kHz. Data displays a pronounced bend after ${\sim}$50s, and a lifetime of $T_2’{=} 357$ s --  representing a ${\sim}$238,000-fold increase over $T_2^{\ast}$. Surprisingly, this large enhancement occurs despite only ${\app}$20 pulses being applied per $T_2^{\ast}$ period. Importantly, each curve in \zfr{mfig1}D is sampled rapidly. Inset \zfr{mfig1}D(i) highlights this, with ${\app}$1300 data points per ${\app}$100 ms window; the red trace itself has ${>}$8M points in total (upper axis).

The marked lifetime estimates above are from $1/e$ intercepts (shown by the dashed line). While common practice, these estimates form an incomplete description (adequate only when decays are monoexponential),  and provide limited insights into decay processes. Instead, leveraging rapid data sampling and high SNR, we employ Laplace Inversion (LI) \cite{song2002t1, venkataramanan2002solving}, decomposing the traces via an exponential kernel, $S{=}\sum_jw_j\exp(-t/T_2'^{j})$ to extract individual weights $w_j$ and $T_2'^{j}$ values (as detailed in SI~\zsr{laplace}). 

Blue trace in \zfr{mfig1}E(i) shows the result for on-resonance data from \zfr{mfig1}D. The high SNR and rapid data sampling enables producing a high-resolution $T_2'$ (LI)  spectrum. Data in \zfr{mfig1}E(i) reveals seven distinct, narrow peaks with $T_2'^{j}$ values at 0.8s, 2.5s, 9s, 28s, 75s, and 170s, respectively, individual groups separated apart about half an order of magnitude (dashed vertical lines). Further details on the factors influencing LI spectral resolution, and effects of noise and data truncation, are discussed in SI~\zsr{laplace}. Red trace (\zfr{mfig1}E(ii)) instead shows the off-resonance case from \zfr{mfig1}D. Short-time $T_2'^{j}$ components remain nearly identical (dashed lines), while the long-time component significantly lengthens, with the longest reaching $T_2'^{j}{\app}$680s.

\zfr{mfig1}F(i) demonstrates the efficacy of the LI fit, using \zfr{mfig1}E(ii) as a representative example. Data points from \zfr{mfig1}D are shown in red on a semi-logarithmic time scale, while the LI derived fit from \zfr{mfig1}E(ii) is in purple. There is an excellent overlap of the fit with the data. Bottom panel (\zfr{mfig1}F(ii)) shows the fit residuals, which remain under 1\% across the entire ${>}$8M point, 600s, dataset capturing well both the short- and long-time dynamics.

\zfr{mfig1}E-\zfr{mfig1}F demonstrate that LI constitutes an efficient approach for ``noise spectroscopy''~\cite{alvarez2011measuring}, revealing the individual components causing $T_2'$ relaxation of the $^{13}$C nuclei. Unlike commonly used dynamical decoupling methods~\cite{viola1999dynamical, alvarez2011measuring,yuge2011measurement,norris2016qubit}, the analysis here is performed from a single experimental shot. The results are interesting because they reveal a series of discrete, well-separated, components spanning about five orders of magnitude.

\subsection{Extended transverse lifetimes by self-decoupling}
\zsl{LG}

To investigate the origin of the surprising extension in \zfr{mfig1}D, \zfr{mfig2} examines the influence of frequency offset over 51 values from $\xD\omega{\in}[-5,5]$ kHz, using pulses with a nominal flip angle $\xt{=}90^\circ$ and performing an LI analysis of the normalized time-traces taken to $t{=}600$s, similar to \zfr{mfig1}E. \zfr{mfig2}A shows the LI results as a 3D plot; $T_2'^{j}$ values are plotted on the logarithmic vertical axis, while weights $w_j$ are represented by colors (see colorbar). The solid green spline curve is a guide to the eye highlighting the long-time $T_2'^{j}$ component. It demonstrates a significant increase at $\xD\omega{\app}\pm$2 kHz (red dashed line) where $T_2'^{j}{\app}680$s, leading to the characteristic horn-like features. In contrast, the shorter-time components remain largely unchanged regardless of offset (gray dashed lines). Black points (joined by the dark gray line) show the corresponding $T_2'$ values obtained from just a $1/e$ intercept. While it displays a similar trend, the information content within it is much lower.

To understand the horn-like features in more detail, \zfr{mfig2}B presents an analogous set of experiments for the right half of \zfr{mfig2}A, i.e., $\xD\omega{\in}[0,5]$ kHz, with varying pulse power and fixed nominal $\xt{=}90^{\circ}$. The data reveals that as the time-averaged (effective) Rabi frequency $\Omega$ increases (moving down in \zfr{mfig2}B), the horn-like feature of the long-time $T_2'^{j}$ component shifts to higher $\xD\omega$, as indicated by the highlighted $\xD\omega$ values and the dashed lines.

In SI \zsr{theory} we construct a minimal theoretical model to explain the observations. The observed shift of the peak $\xD\omega$ value is in good qualitative agreement with this analysis, and suggests that the $T_2'$ extension results from matching the Lee-Goldburg condition~\cite{lee1965nuclear,mehring1972magic}, where at $\xD\xo\approx \Omega/\sqrt{2}$, the effective spin-locking axis aligns at the magic angle, decoupling the $\Cs$ nuclei~\cite{haeberlen1968coherent} (see SI ~\zsr{theory}). This yields a time-averaged effective Hamiltonian $\mHeff{=}0$, that vanishes to zeroth order in the Magnus expansion, \I{self-decoupling} the $\Cs$ nuclei. On-resonance, however, $[\mHeff,\xr_0]{=}0$. \zfr{mfig2}C schematically illustrates both cases. While both lead to an extended $T_2'$, we rationalize the extension in the LG case as due to (1) the reduced weight of higher-order Magnus terms (see SI~\zsr{theory}), and (2) the suppression of spin-diffusion effects among $\Cs$ nuclei. In contrast, spin diffusion remains active in the resonant case, where lattice electronic spins (NV and P1 centers) open channels of $\Cs$ relaxation via those nuclei proximal to them~\cite{beatrez2023electron}.

\subsection{Anomalously long lifetimes under small angles}
\zsl{small-angle}
While \zfr{mfig2} considered the case of $\xt$ nominally $\xt{=}90^{\circ}$, \zfr{mfig3} instead explores the impact of \I{small-angle} pulses applied off-resonance. \zfr{mfig3}A shows normalized decay traces on a semi-logarithmic scale with nominal $\xt{=}5^\circ$ and $\qt$ identical to \zfr{mfig2}A and 100K; colors represent different offset values $\xD\xo$. Since the time-averaged Rabi frequency is significantly lower than in \zfr{mfig1}D and \zfr{mfig2}A, the $T_2'$ lifetimes are expected to decrease~\cite{abanin2015exponentially,beatrez2021floquet}. Indeed, this is borne out in the resonant case, where $T_2'{=}34$s (bottom trace in \zfr{mfig3}A).

Surprisingly, however, we observe a dramatic increase in $T_2'$ is upon going off-resonance (see \zfr{mfig3}A). At $\xD\xo{=}5$kHz -- corresponding to applying pulses far into the wing of the $\Cs$ spectrum -- we estimate $T_2'{\app}800$s (purple curve in \zfr{mfig3}A) from the $1/e$-crossing. This corresponds to a remarkable ${>}$533,000-fold increase in $T_2'$ lifetime over the free induction decay $T_2^*$ value, and constitues a record value for extension comparing other platforms where Floquet prethermalization has been observed before (see comparison table in \zfr{SItable}). Importantly, this is achieved with a rather minimal amount of spin control: within one $T_2^*$ period, the pulses only add up to a single (nominal) $\pi$ rotation. We therefore term this an \I{anomalous} extension in the $T_2'$ lifetime.

Another noteworthy finding is that the traces in \zfr{mfig3}A become progressively more monoexponential with higher $\xD\xo$. To illustrate this, \zfr{mfig3}B shows a $T_2'$ LI map of the normalized traces, similar to \zfr{mfig2}A, as a function of $\xD\xo$. The data reveals a significant increase in the position and weight of the long-time $T_2'$ component away from resonance, while the shorter-time components diminish. The cyan solid line is a spline fit to the longest component, while black points show the corresponding $1/e$ intercepts. \zfr{mfig3}C presents line cuts at $\xD\xo{=}0$ (shaded) and 5 kHz (blue); for the latter, the $T_2'$ spectrum is dominated by a single component at $T_2'{\app}800$s estimated by the centroid.

\zfr{mfig3}D now examines pulse angle effects by varying nominal pulse angle $\xt$ with fixed $\qt$, showing $T_2'$ LI maps across three resonance offsets: (i) on-resonance, (ii) $\xD\xo{=}2$ kHz, and (iii) $\xD\xo{=}5$ kHz. A fuller description of the data is in SI \zsr{pulse-width}. For the on-resonance case (\zfr{mfig3}D(i)), we observe that smaller $\xt$ values reduce the $T_2'$ lifetimes. This is seen in the downward trend of the long-time $T_2'$ component (marked by the cyan line), as well as that from a simple $1/e$ intercept (black points). This behavior is along expected lines: smaller $\xt$ pulses yield a reduction in time-averaged Rabi frequency, and lower prethermal lifetimes~\cite{beatrez2021floquet}. 

In contrast, however, at $\xD\xo{=}5$kHz (\zfr{mfig3}D(iii)), the long-time $T_2'$ component (cyan line) bends upward at small $\xt$, a reflection of the anamalous behavior. The $\xD\xo{=}2$kHz case (\zfr{mfig3}D(ii)) falls in-between. Analogous data to \zfr{mfig3}D instead showing variations with changing pulsing frequency ($\qt^{-1}$) is discussed in SI \zsr{pulse-frequency}. 

The theoretical model in SI \zsr{theory} is able to at least qualitatively explain these observations. In short, the prethermal axis aligns more towards $\zhat$ with increasing offset, schematically shown in \zfr{mfig3}E, allowing $T_2'$ to asymptotically approach the $T_1$ limit far off-resonance (see SI \zsr{theory}). 

We emphasize finally that the data in \zfr{mfig3}A refer to normalized traces. Tilting of the prethermal axis towards $\zhat$ with increasing $\xD\xo$ leads to a trivial loss in absolute signal. For clarity, \zfr{mfig3}F shows the initial signal amplitude (at $t{=}0$) as a function of $\xD\xo$ for nominal $\xt{=}5^\circ$. However, this does not impact sensing, as the focus there is on signal \I{deviations} $\xD S$ (see \zsr{sensing}).

\subsection{Application: Long-time continuously interrogated AC magnetometry}
\zsl{sensing}

While already fundamentally interesting, the extended prethermal lifetimes can be also exploited for quantum sensing. We consider, in particular, the sensing of time-varying (AC) magnetic fields $\Bsens(t)=\Bac\cos(2\pi\fac t+\xph)$ using the hyperpolarized $\Cs$ nuclei~\cite{sahin2022high}. \zfr{mfig4}A shows the protocol, here at room temperature (RT); the $\Cs$ spins are exposed to the AC field along $\zhat$ with the Floquet sequence from \zfr{mfig1}C, while being quasi-continuously interrogated. The AC field does not need to be synchronized with the sequence. The sensing principle, previously described~\cite{sahin2022high}, involves $\Cs$ nuclei prethermalizing to a time-varying Hamiltonian that includes the AC field; the spin micromotion dynamics then reports a direct imprint of $\Bsens(t)$. 
This results in oscillatory deviations $\xD S$ on top of the signal $S$ in \zfr{mfig1}D and \zfr{mfig3}A, with $\xD S \propto \Bac$, and frequency at harmonics of $\fac$ (dominantly $\fac$ or $2\fac$).

Sensing is effective over a relatively wide bandwidth $\fac\in[$10 Hz,1 MHz$]$; the upper limit determined by the shortest interval between pulses. This sensing bandwidth lies in a blind-spot~\cite{moon2024DTC} between vapor cell magnetometers~\cite{budker2007optical,ijsselsteijn2012full,kurian2023single} and electron-based sensors such as NV centers \cite{mazeNanoscaleMagneticSensing2008, kuwahata2020magnetometer, sahin2022high}, and is therefore in a practical useful range.

\zfr{mfig4}B shows the RT sensing result for a single-tone AC field at $\fac{=}50$Hz with $\Bac{=}82\mu$T. In contrast to other quantum sensing platforms~\cite{degen2017quantum}, only a single-shot of initialization (hyperpolarization) is applied here.  Signal deviations $\xD S$ are then measured continuously, without sensor reinitialization, for $t{\sim}10$min. For clarity, data in \zfr{mfig4}B shows $\xD t{=}100$ ms windows at intervals of 1min, 3min, 5min, and 9.6min respectively. Two cases are compared: \zfr{mfig4}B(i) shows on-resonant pulses with flip-angle $\xt{=}90^{\circ}$ (green), and \zfr{mfig4}B(ii) shows $\xD\xo{=}5$kHz off-resonant pulses with nominal $\xt{=}10^{\circ}$ (yellow). As is evident, $\xD S$ reflects a direct imprint of $\Bac$; the relative weights of the $\fac$ and 2$\fac$ components depends on the offset. In \zfr{mfig4}B, data is shown by the light lines, while the darker lines are a fit. The numbers marked in grey indicate the measured SNR of the AC field signature over the noise floor in a $1$s interval for each window. 

Physically, for sensing, one is interested in the excursions of the spin vector away from the originally defined prethermal axis due to the AC field; measuring its projection on the $\xy$ plane. While the on-resonance signal $S$ might have a large absolute (DC) magnitude (not shown in \zfr{mfig4}B, see \zfr{mfig3}F), this is not relevant for sensing. Instead, the strength of the oscillations riding atop it that constitute the sensing signal $\xD S$ is, which is in fact larger in the off-resonance case. In addition, the on-resonance $\xD S$ signal decays
significantly faster, as is evident moving rightwards in \zfr{mfig4}B. By $t{=}580$s, the off-resonance, small-angle, signal (\zfr{mfig4}B(ii)) shows an SNR that is almost an order of magnitude larger than the on-resonance case (\zfr{mfig4}B(i)).

For a more comprehensive view into the sensing gains using small-angle off-resonance pulses, \zfr{mfig4}C-D contrasts $\xt{=}90^\circ$ (green) and $\xt{=}15^\circ$ (orange), showing the $\xD S$ signal intensity at $\fac{=}50$Hz harmonics as a function of offset $\xD \xo$. Two cases are considered: integrated $\xD S$ signal, denoted $\int\xD S$ for simplicity, over a $t{=}$3min experiment (\zfr{mfig4}C), and over a $\xD t{=}1$s window at $t{=}170$s (\zfr{mfig4}D). In both panels, the shaded region indicates the detection noise floor, determined just by cavity Johnson noise. Notably, we observe that the small-angle case outperforms across the entire range; even on-resonance ($\xD \xo{=}0$), it yields about a 10-fold greater sensitivity. At $\xD \xo{=}5$kHz off-resonance, the $\xt{=}90^\circ$ signal (green traces) decays rapidly while the small-angle signal does not. 

\zfr{mfig4}E shows $T_2’$ values estimated from $1/e$ intercepts for both cases at room temperature, closely matching trends in \zfr{mfig2}A and \zfr{mfig3}B. We identify, for instance, the horn-like feature around $\xD\xo{=}2$kHz in the green trace; and the lower $T_2'$ lifetime for the small-angle case on-resonance which rapidly increases upon increasing $\xD\xo$.

Finally, we comment that while \zfr{mfig4}B considered a single AC tone, any complicated signal within the detection bandwidth will show a similar imprint~\cite{sahin2022high}. We also note that the data in \zfr{mfig4}B has not been optimized for sensitivity, and significant gains are possible through better hyperpolarization~\cite{sarkarRapidlyEnhancedSpinPolarization2022}, and increasing sample occupation (filling-factor) in the readout cavity. At current levels, for resonant fields, we anticipate a comparable time-normalized performance (${\sim}$800pT/vHz) to that reported earlier~\cite{sahin2022high}. However, the ability for long-time sensing demonstrated in \zfr{mfig4}B allows a reduced smallest detectable field via averaging~\cite{rotem2019limits}, and a high-resolution discrimination of AC fields of comparable frequency~\cite{schmitt2021optimal}.

\tocless\section{Discussion}{sec:discussion}
This work demonstrates long-time, reinitialization-free quantum sensing by exploiting extended prethermal lifetimes with small-angle off-resonance pulses. The absolute signal level $S$ of the prethermal state, so long as it is above the noise floor, does not affect the sensing, allowing full exploitation of the long $T_2’$ times (\zfr{mfig4}C-D). The achieved $T_2’{\app}800$s values themselves, asymptotically approaching $T_1$, are significant, and constitute a record value (see \zfr{SItable}). Moreover, experiments in \zfr{mfig4} show sensing for ${\sim}10$min with no re-initialization, also setting a record in both duration and the number of pulses (${\sim}$7M) employed. Even in its current form, this opens applications in sensitive detection in a blind-spot region of ULF and VLF fields~\cite{moon2024DTC} and for underwater magnetometry~\cite{Page21}.

We envision several promising future directions. Further $T_2’$ extension is expected by lowering the temperature to 4K, where at $B_0{=}10$T, NV and P1 center electrons are polarized, suppressing electron-mediated relaxation to the $\Cs$ nuclei~\cite{takahashi2008quenching,jarmola2012temperature}. The protocols developed here are more broadly applicable than just $\Cs$ spins in diamond. We envision similar approaches being applied to ensembles of electronic spins, e.g. NV or P1 centers~\cite{williams2023quantifying}. Particularly compelling may be applications to polarized \(^{1}\)H nuclear spins in organic solids, especially pentacene-doped naphthalene, where long $T_1$ times, due to diamagnetic singlet ground states, can exceed 100hr~\cite{eichhorn2014proton, quan2019polarization,singhRoomtemperatureQuantumSensing2024}. We anticipate long-time, reinitialization-free quantum sensing for several tens of hours in this system.  This portends new applications for nuclear-spin clocks \cite{hodges2013timekeeping}, gyroscopes \cite{ajoy2012stable, ledbetter2012gyroscopes, jaskula2019cross, jarmola2021demonstration}, relayed NMR sensors for proximal analytes \cite{meriles2005optically,sidles2009spin}, and spin-based magnetometers akin to levitating compass needles \cite{jackson2016precessing}, which can be continuously responsive to external magnetic fields, while capable of being tracked in three-dimensions on the Bloch sphere \cite{sahin2022continuously}.

The Laplace inversion approach introduced here offers a novel means of discerning individual mechanisms for nuclear spin relaxation. We observe that the Laplace peaks are organized into discrete peaks (\zfr{mfig1}E, \zfr{mfig2}A) likely reflecting different mechanisms of nuclear decay, such as dipolar coupling, electron-nuclear interactions, and phonon processes \cite{jarmolaTemperatureMagneticFieldDependentLongitudinal2012, beatrezElectronInducedNanoscale2023}. The transition to mono-exponential behavior via off-resonant excitation (\zfr{mfig3}C) suggests the ability to tailor the spectral density profiles experienced by the nuclei \cite{ajoyHyperpolarizedRelaxometryBased2019}.  This anticipates future work controlling individual decay channels via quantum control or changes in temperature or magnetic field \cite{takahashi2008quenching}.

\paragraph*{Acknowledgements.}
We gratefully acknowledge discussions with M. Bukov, V. Ivanov, P. Schindler, C. Ramanathan, and L. Tan. This work was funded by DNN NNSA (FY24-LB-PD3Ta-P38), ONR (N00014-20-1-2806), AFOSR YIP (FA9550-23-1-0106), AFOSR DURIP (FA9550-22-1-0156), and the CIFAR Azrieli Foundation (GS23-013). YS acknowledges the support by ARPA-E (Program Directors, Dr. Isik Kizilyalli, Dr. Olga Spahn, and Dr. Doug Wicks, under Contracts DE-AR0001063 and DE-AR0001705). KAH acknowledges an NSF Graduate Research Fellowship. CS acknowledges an ARCS Graduate Fellowship.

\paragraph*{Author contributions.}
{\bf Kieren Harkins:} Investigation; instrument development; data analysis and writing – original draft. {\bf Cooper Selco:} Investigation; data analysis and writing – original draft. {\bf Christian Bengs:} Formal analysis; software development and writing – original draft. {\bf David Marchiori:} Investigation; instrument development. {\bf Leo Joon Il Moon:} Investigation; data analysis. {\bf Zhuo-Rui Zhang:} Data curation and visualization. {\bf Aristotle Yang:} Data validation and visualization. {\bf Angad Singh:} Data validation and visualization. {\bf Emanuel Druga:} Instrument development. {\bf Yi-Qiao Song:} Supervision; software and formal analysis. {\bf Ashok Ajoy:} Project supervision; conceptualization; formal analysis and writing – original draft.

\bibliography{TOFPaper2}  % Produces the bibliography via BibTeX.

%\clearpage
%\appendix

%%%%%%%%%%%%%%%%%%%%%%%%%%%%%%%%%%%%%%%%%%%%%%%%%%%%%%%%%%%%%%%%%%%%%%%%%%%
%                        SI
%%%%%%%%%%%%%%%%%%%%%%%%%%%%%%%%%%%%%%%%%%%%%%%%%%%%%%%%%%%%%%%%%%%%%%%%%%%
%%%%%%%%%%%%%%%%%%%%%%%%%%%%%%%%%%%%%%%%%%%%%%%%%%%%%%%%%%%%%%%%%%%%%%%%%%%

\clearpage
\onecolumngrid
%\begin{widetext}
\begin{center}
\textbf{\large{\textit{Supplemental Information:} \\ \smallskip Anomalously extended Floquet prethermal lifetimes and applications to long-time quantum sensing}} \\\smallskip
\end{center}

\twocolumngrid

\beginsupplement
\tableofcontents

%%%%%%%%%%%%%%%%%%%%%%%%%%%%%%%
%            Text
%%%%%%%%%%%%%%%%%%%%%%%%%%%%%%%
%%%%%%%%%%%%%%%%%%%%%%%%%%%%%%%

\begin{figure*}[t]
  \centering
 {\includegraphics[width=0.85\textwidth,trim={1cm 7.1cm 0.5cm 2.5cm},clip]{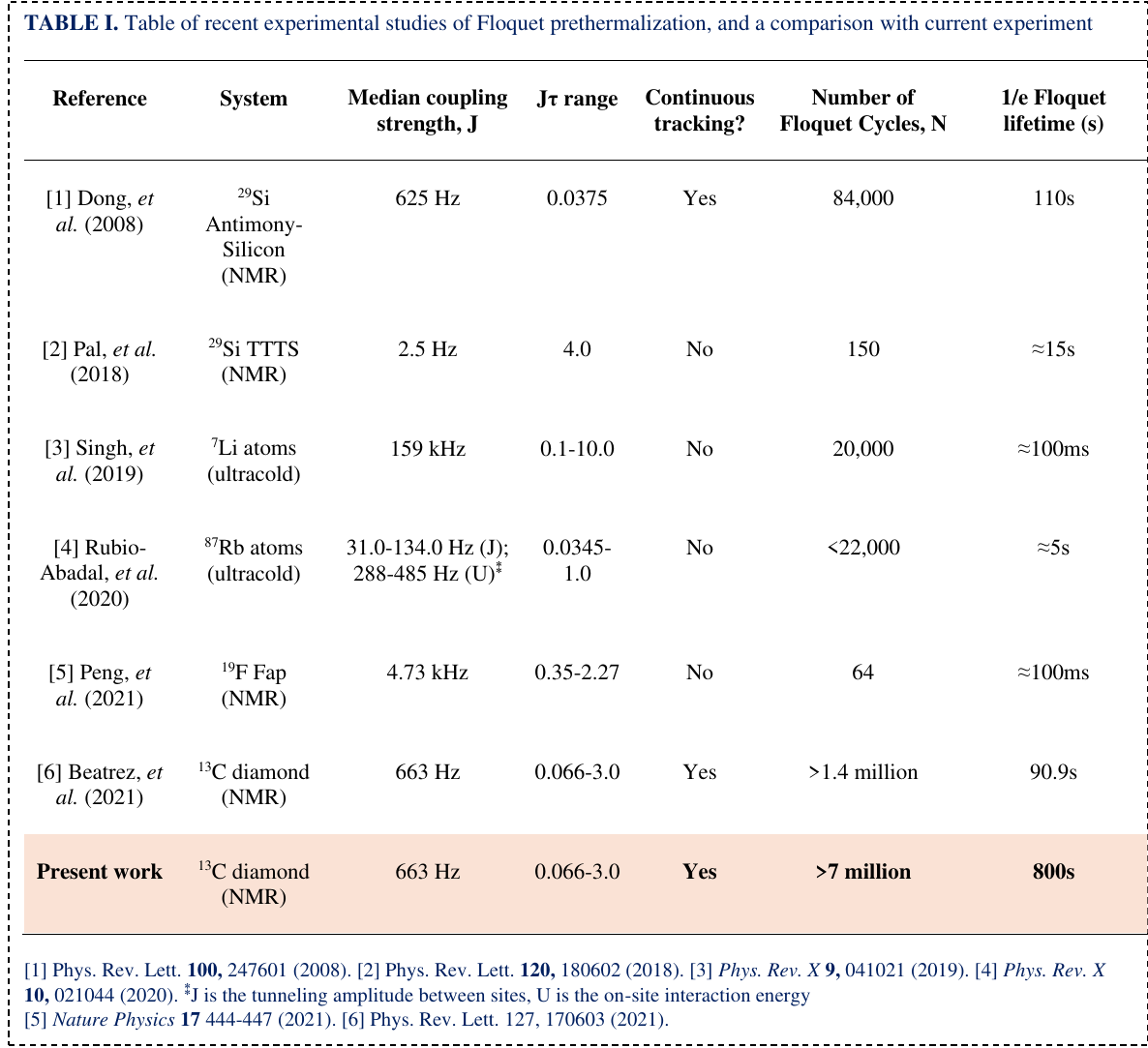}}
    \caption{\T{Table of recent experimental studies} of Floquet prethermalization, and a comparison with current experiment. Floquet cycles are defined as the periodicity of the basic pulse building block.
}
\zfl{SItable}
\end{figure*}

\begin{figure*}[t]
  \centering
 {\includegraphics[width=0.85\textwidth]{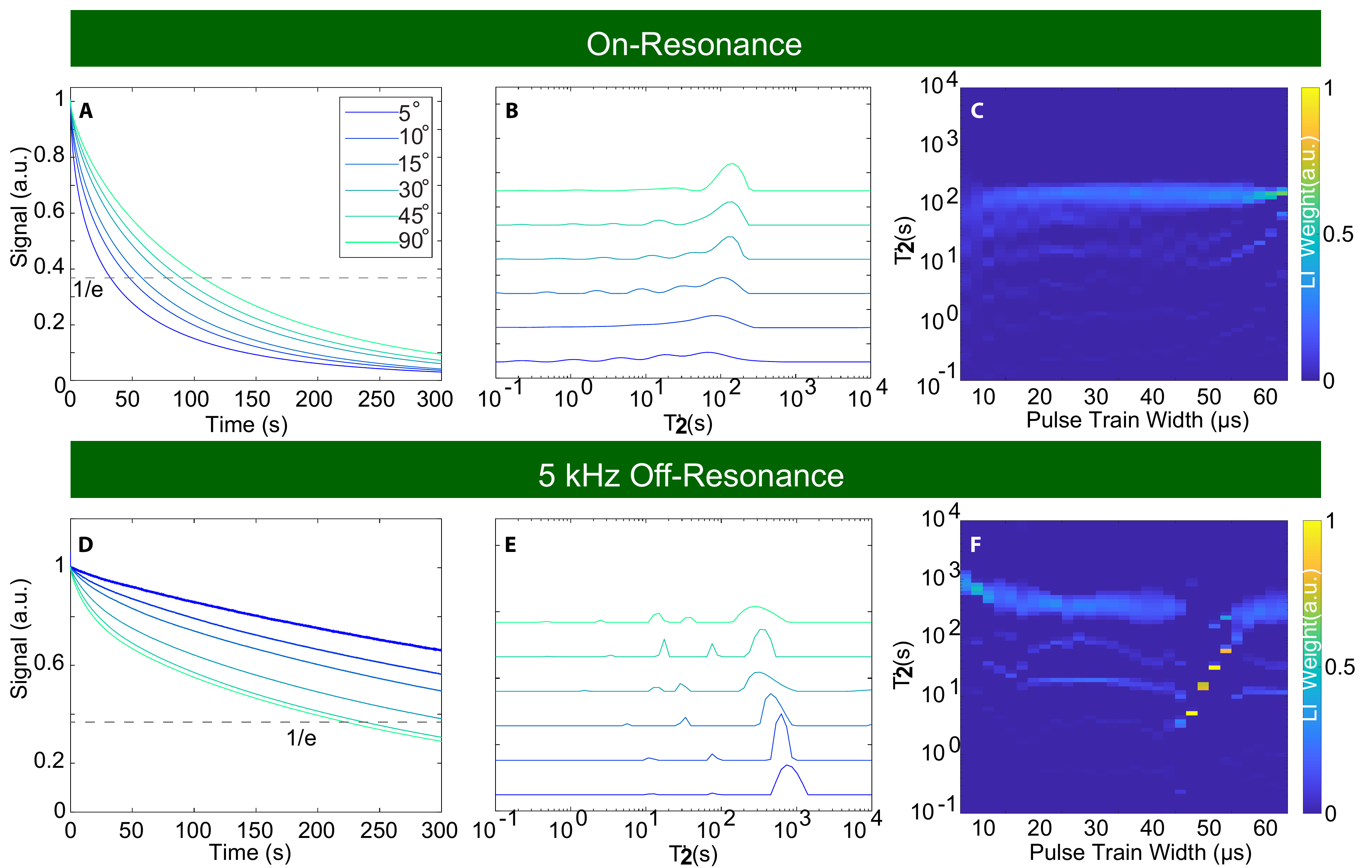}}
    \caption{\T{Variation of $T_2’$ decay times with pulse width} ($t_p$) for two cases: (A-C) On-resonance and (D-F) 5 kHz off-resonance.
(A) Representative traces show spin-lock decay with interpulse spacing $\tau = 43 \mu s$ and varying pulse width. $T_2’$ lifetime decreases with decreasing $t_p$ on-resonance. 
(B) LI traces show lengthening of the dominant $T_2’$ component with increasing pulse width.
(C) Color plot of the LI map for 40 slices of $t_p$ values. Vertical axis shows individual $T_2’$ components on a log scale, horizontal axis shows varying $t_p$ values (2-34 $\mu$s), corresponding to flip angle $\xt$ change (5-90 degrees). Colors represent LI intensity of the normalized curves. Low-$T_2’$ components show negligible change with increasing pulse width; long-time component increases slightly. 
(D-F) Same plots for data at 5 kHz off-resonance. 
(D) Lengthening of decay curves with decreasing $t_p$ is evident. 
(E) LI traces show a significant increase in $T_2’$ components with decreasing $t_p$, revealing an increasingly mono-exponential character at small $t_p$.
(F) LI map highlights a significant increase in the long-time $T_2’$ component at short $t_p$. Sharp drops at specific pulse widths, around 40 $\mu$s and 70 $\mu$s, correspond to effective $\pi$ rotations during these periods and formation of spin-shell-like textures (see Ref. ~\cite{harkinsNanoscaleEngineeringDynamical2023}).
}
\zfl{SIfig1}
\end{figure*}

\begin{figure*}[t]
  \centering
 {\includegraphics[width=0.85\textwidth]{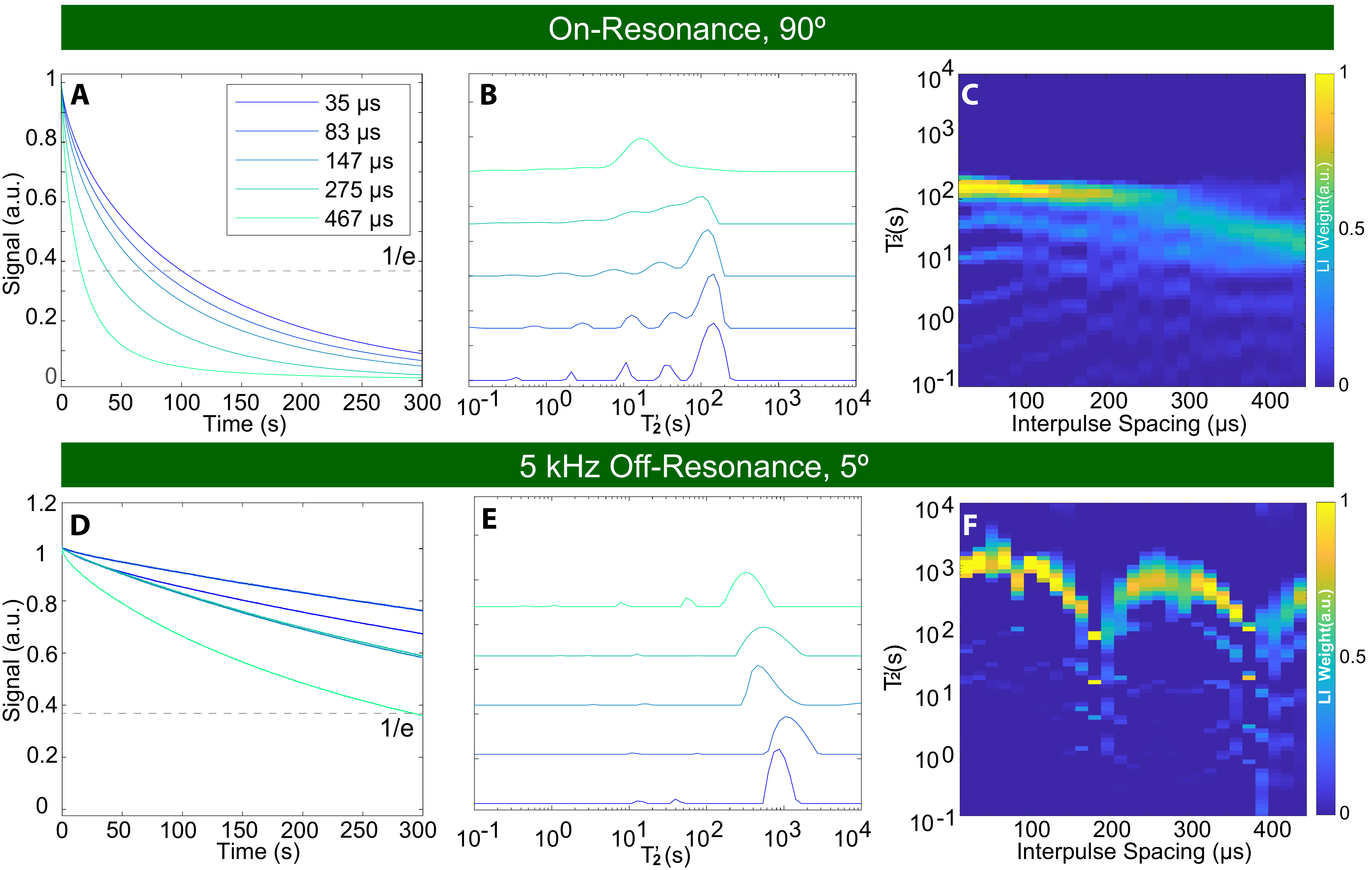}}
    \caption{\T{Variation of $T_2’$ as a function of interpulse spacing $\tau$} under representative conditions (A-B) on-resonance and (C-D) 5 kHz off-resonance.
(A) \I{On-resonance traces} for varying $\tau$ (see legend) with a fixed 90-degree pulse. Increasing pulsing frequency (decreasing $\tau$) leads to an increase in $T_2’$ lifetimes.
(B) \I{LI maps} for 28 slices with $T_2’$ values on the vertical axis (logarithmic scale) and interpulse delay $\tau$ in $\mu$s on the horizontal axis. Lengthening of $T_2’$ values at short $\tau$ is indicated by the slight negative slope in the traces.
(D) \I{Off-resonance data} at 5 kHz offset with $\theta = 5^\circ$ shows a similar trend but with a significant increase in lifetime. $T_2’$ lifetime here exceeds 500 s at short $\tau$ values.
(E) \I{LI maps} indicate that at short interpulse delays (here for $\theta = 5^\circ$), $T_2’$ lifetime exceeds 1000 s. Sharp dips correspond to matching a $\pi$ condition and the formation of stable spin shells (Ref. ~\cite{harkinsNanoscaleEngineeringDynamical2023}).}
\zfl{SIfig2}
\end{figure*}

\begin{figure}[t]
  \centering
 {\includegraphics[width=0.5\textwidth]{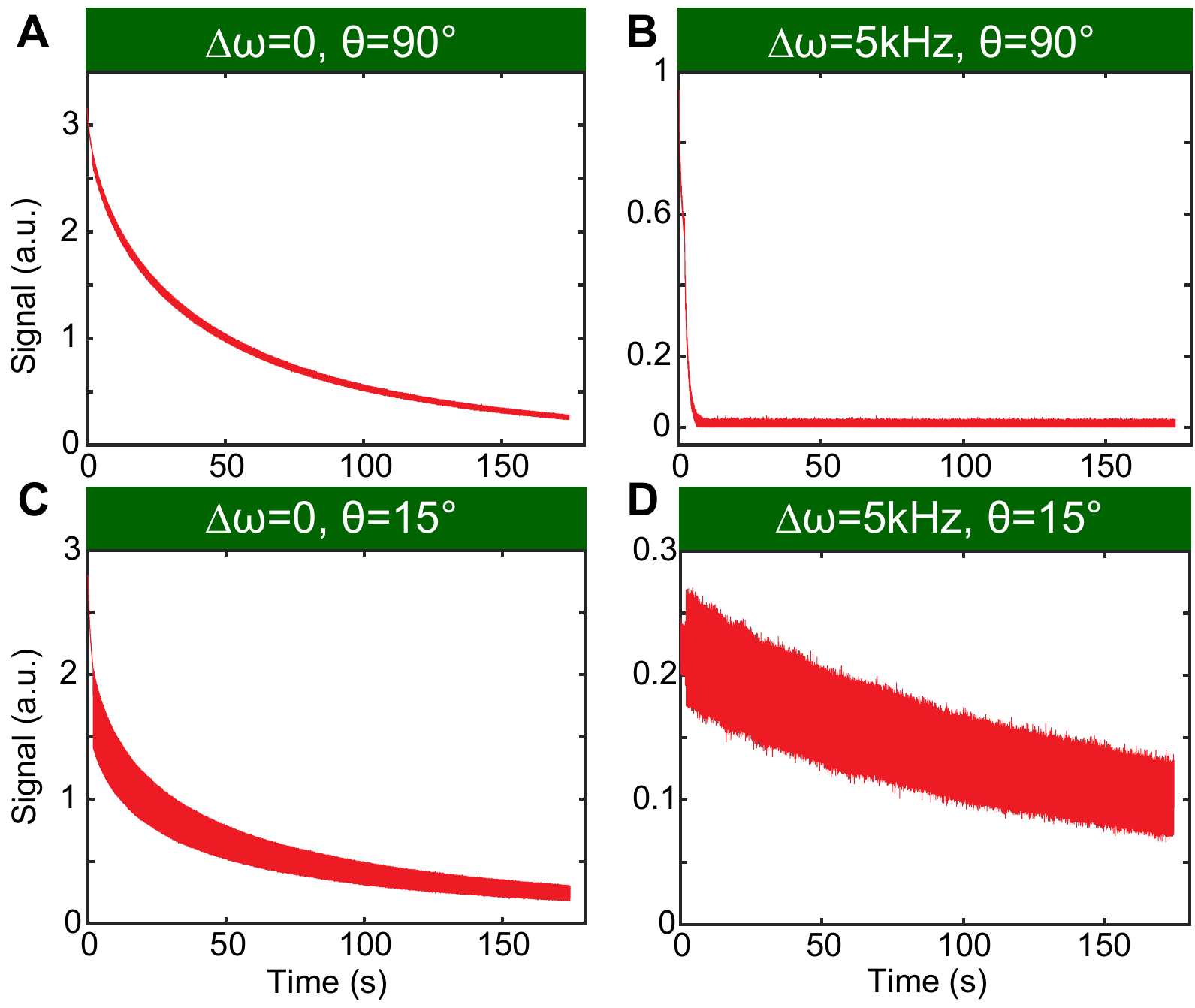}}
    \caption{\T{Complete time domain profiles of long-time sensing experiments}. Each of the four profiles shows the extremes of the $\app$180s data which was used in \zfr{mfig4}C-E. (A) Offset $\Delta\omega {=} 0$ and nominal flip angle $\xt {=} 90^{\circ}$, corresponding to the left-most green point in \zfr{mfig4}C-E. (B) Offset $\Delta\omega {=} 5$kHz and nominal flip angle $\xt {=} 90^{\circ}$, corresponding to the right-most green point in \zfr{mfig4}C-E. The signal decays extremely rapidly under the influence of the external AC field, explaining the loss in sensitivity $\int\Delta$S and the very short lifetime $T_{2}'$. (C) Offset $\Delta\omega {=} 0$ and nominal flip angle $\xt {=} 15^{\circ}$, corresponding to the left-most orange point in \zfr{mfig4}C-E. The initial decay is rapid (explaining the short lifetime $T_{2}'$), but there is a long-lived component which is sensitive to the external AC field, explaining the high sensitivity $\int\Delta$S. (D) Offset $\Delta\omega {=} 5$kHz and nominal flip angle $\xt = 15^{\circ}$, corresponding to the right-most orange point in \zfr{mfig4}C-E. The initial loss in signal is large, but the lifetime $T_{2}'$ is very long (see \zfr{mfig3}(B) and \zfr{mfig4}E), leading to a high sensitivity to the external field. 
}
\zfl{SIfig3}
\end{figure}

\begin{figure}[t]
  \centering
 {\includegraphics[width=0.5\textwidth]{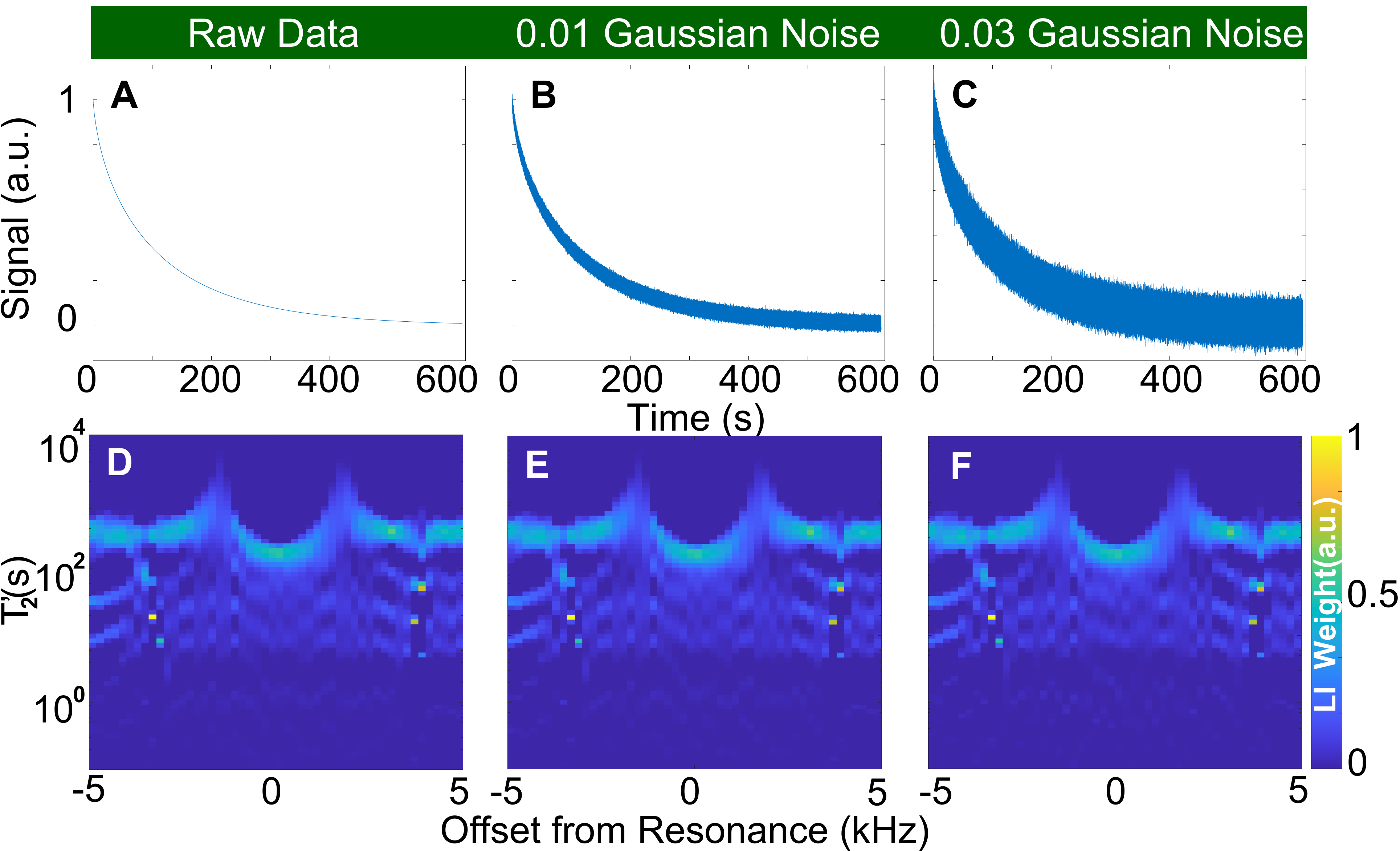}}
    \caption{\T{Robustness of LI fits against noise.} 
    (A)-(C) \I{On-resonance slice of data with different amounts of Gaussian noise added to it.} Data is taken beyond $t=600$s and sampled every $\tau = 43$ microseconds, yielding $>8$M data points. (A) shows the raw data, while (B)-(C) show data with white Gaussian noise of amplitude 0.01 and 0.03 added to the raw data, respectively. 
    (D)-(F) \I{LI maps} using the same data in \zfr{mfig2}A of the main text with different amounts of Gaussian noise added to them (0, 0.01, and 0.03 corresponding to panels (A)-(C)). It can be seen that the LI algorithm is robust even against large amounts of noise added to the raw data as the LI map has almost no change across the three different noise levels.}
\zfl{SIfig4}
\end{figure}

\begin{figure}[t]
  \centering
 {\includegraphics[width=0.5\textwidth]{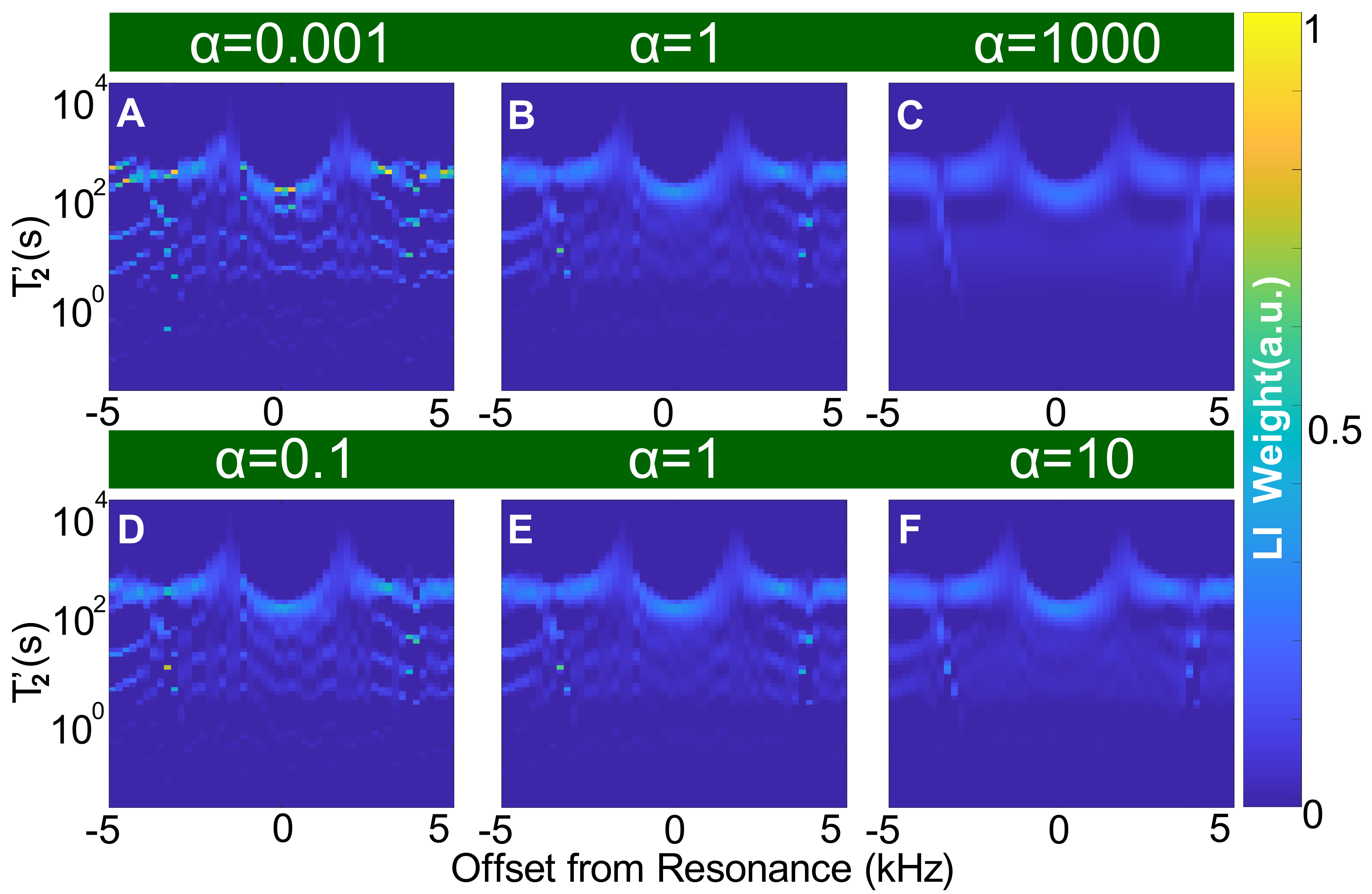}}
    \caption{\T{Impact of regularization parameter $\alpha$ on the LI fits.} Panels show LI colormaps using the same data as \zfr{mfig2}A of the main paper with five values of $\alpha$: (A) $10^{-3}$, (B) 1 (C) 1000, (D) 0.1, (E) 1, and (F) 10.
(A) \I{Small $\alpha$} results in higher resolution LI fits but risks potential overfitting, especially at shorter $T_2'^{j}$ components.
(B) Compromise value of $\alpha = 1$ balances resolution and overfitting.
(C) \I{Large $\alpha$} causes LI components to blur, reducing effective resolution.
(D)-(F) Varying $\alpha$ over 1-2 orders of magnitude does not significantly change the results of the LI fit, however, varying $\alpha$ over many orders of magnitude ((A)-(C)) can have significant changes on the outcome of the LI fit.
}
\zfl{SIfig5}
\end{figure}

\begin{figure}[t]
  \centering
 {\includegraphics[width=0.5\textwidth]{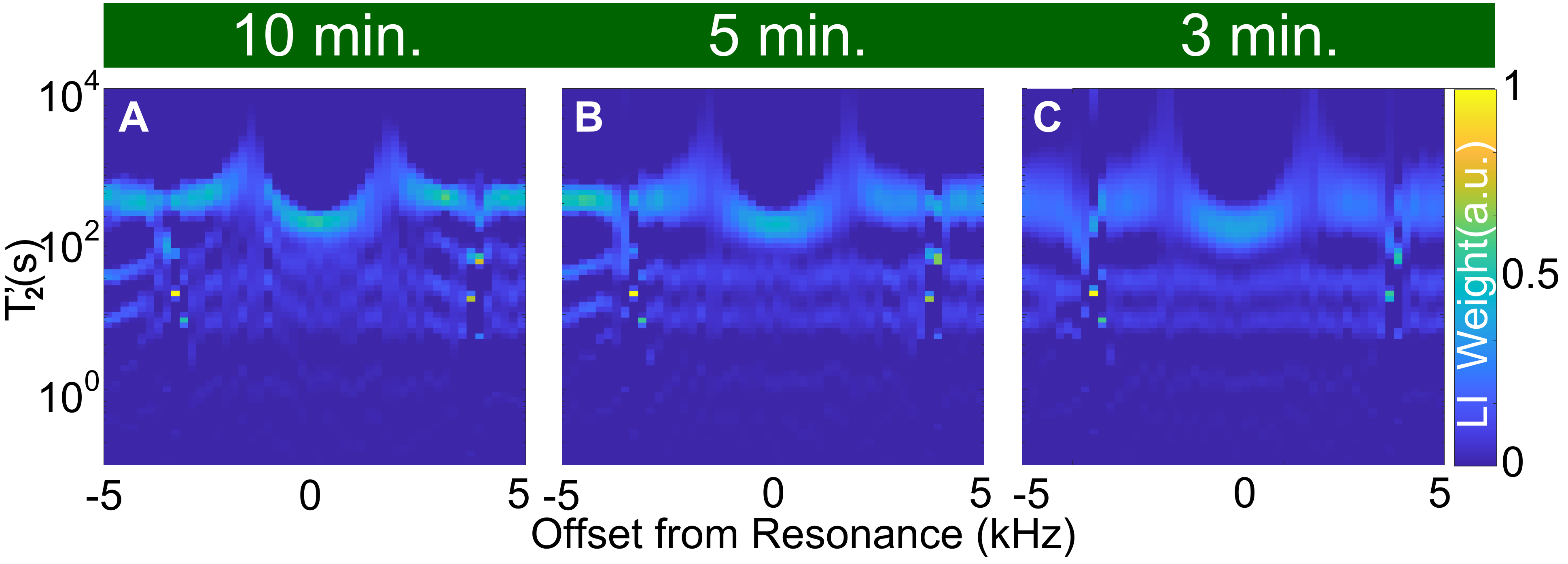}}
    \caption{\T{Effect of truncating data on LI maps}. Shown are three cases: data measured for (A) 10 min (closer to the ground truth with complete signal decay) and data truncated at (B) 5 min and (C) 3 min. 
(A) \I{Fitting quality} is best for the 10-minute data where the entire decay trace is captured. 
(B-C) \I{Truncated data} shows similar behavior, allowing technically sound conclusions about individual $T_2’^{j}$ components despite the truncation, however, some of the finer details of the LI fit are lost.
}
\zfl{SIfig6}
\end{figure}

\section{Materials and Methods}
\T{Materials -- } The sample used in this work is a single-crystal diamond measuring 3 ${\zt}$ 3 ${\zt}$ 0.3 mm, containing a natural abundance of $\Cs$ nuclei and NV centers at ~1 ppm concentration. This same sample has been characterized in prior studies on Floquet prethermalization~\cite{beatrez2021floquet}, allowing for direct comparisons to the lifetime extensions observed. The sample is oriented parallel to the \(B_{\text{pol}} \app 27\) mT magnetic field, ensuring simultaneous hyperpolarization of the four NV center axes.

\T{Experimental Setup -- } Instrumentation for hyperpolarization and $\Cs$ readout follows previous works, and we refer the reader to those studies for detailed descriptions~\cite{ajoy2018orientation}. Data here is taken with two apparatus: one at 100K and 9.4T (\zfr{mfig1}-\zfr{mfig3}) and another at room temperature and 7T (\zfr{mfig4}). Polarization in both cases is carried out at low fields, \(B_{\text{pol}} \approx 30\) mT, driven by optically excited NV centers and chirped microwave (MW) excitation. The polarization mechanism involves successive Landau-Zeener anti-crossings in the rotating frame \cite{ajoy2018orientation,ajoy2021low, pillai2023electron}.

The low-temperature DNP apparatus uses a shuttled cryostat design described in Ref.~\cite{harkins2023nanoscale}. The sample is polarized in a cryostat at the \(B_{\text{pol}}\) center, located in the fringe field above the \(B_0 {=} 9.4\) T superconducting magnet, with a laser impinging from below. The cryostat is then shuttled to \(B_0\) in ${\sim}$90s for NMR measurements. This cryostat hosts a dual NMR-hyperpolarization probe with a MW coil for DNP and an NMR saddle coil for inductive readout of the $\Cs$ precession signal.

\T{$\Cs$ spin control and readout -- } $\Cs$ interrogation is performed using a high-speed arbitrary waveform transceiver (Proteus), as described in Ref.~\cite{moon2024experimental}. This device features a high memory capacity (16GB) and a large sampling rate (1GS/s), enabling continuous interrogation of $\Cs$ spin precession in windows between pulses. In typical experiments, we apply 4-12M pulses. The readout window is typically ${\app}16\mu$s, but for long acquisitions (${>}$300s), we shorten it to ${\app}4\mu$s to conserve memory.

The entire Larmor precession is sampled in these windows, allowing us to discern both amplitude and phase information of the spins in the rotating frame. This is particularly useful in magnetometry experiments where the signal of interest is imprinted in both amplitude and phase. For AC magnetometry, the spins are exposed to a weak magnetic field applied via a secondary coil, parallel to the microwave excitation loop and positioned within the NMR probe. The field strength is calibrated by observing the induced shift in the $^{13}$C Larmor frequency when a DC field is applied of different strengths.

\section{Comparison of Lifetimes to Other Studies of Floquet Prethermalization}
\zsl{Comparison}
\zfr{SItable} shows a table of recent experimental studies of long-time Floquet prethermalization covering a variety of different methods and materials (to the best of our knowledge). For a fair comparison, we refer to a Floquet cycle here as being a basic pulse building block (here of total period $\tau$). For each case lifetimes were approximately determined by the 1/$e$ crossing of the observable under investigation. The bare lifetimes $T_2^*$ without prethermalization are approximately determined by $J^{-1}$ for most platforms. 

As is evident, our results \I{by far} exceed any previous study in terms of the number of applied Floquet cycles and the observed 1/$e$ lifetimes. \zfr{SItable} also displays the corresponding $J\tau$ values for each experiment --  in effect showing a normalized view into how fast the pulses are applied with respect to the inter-spin interaction strength. The extensions obtained in our case (${\app}$533,000-fold) is significantly larger inspite of the pulses being applied at a comparable rate.

\section{Extended data}
\subsection{Effect of Changing Pulse Width on Prethermal Lifetimes}
\zsl{pulse-width}
In \zfr{SIfig1}, we present a detailed analysis of the data from \zfr{mfig3}D of the main paper, examining the effect of pulse width on prethermal lifetimes under different offset conditions. We consider the on-resonance case and the 5 kHz off-resonance case.

For the on-resonance setting \zfr{SIfig1}A-C, we observe an increase in lifetime with increasing pulse width. This can be understood as arising from a larger time-averaged Rabi frequency $\Omega$ at larger pulse widths, and holds for all pulse widths except those with flip angle $\xt\app\pi$, where more complex spin dynamics result, forming shell-like spin textures within the nuclear spins \cite{harkins2023nanoscale}. 

However, this picture is inverted off-resonance \zfr{SIfig1}D-F. At 5 kHz off-resonance, normalized traces show a significant increase in lifetime for shorter pulses (see \zfr{mfig3}D(iii) color plot of the main paper). Moreover, lifetimes off-resonance exceed those on-resonance. This novel, anomalous, lifetime extension allows long-period sensing in a continuously interrogated, reinitialization-free, manner.

The sharp dips in \zfr{SIfig1}F correspond to satisfying an effective \(\pi\) condition, where a stable shell-like spin texture forms in the $\Cs$ spins. For more details, we refer readers to Ref.~\cite{harkins2023nanoscale}. Essentially, in the presence of the gradient magnetic field from the NV center, there is an effective inversion of magnetization in a spatially dependent manner, forming a shell-like texture with two opposite signs of magnetization on either side, and which can remain stable for long periods.

\subsection{Effect of Changing Pulsing Frequency}
\zsl{pulse-frequency}
In a complementary set of experiments (\zfr{SIfig2}), we consider the effect of changing the pulsing frequency while keeping the pulse width fixed under different resonance offset conditions by varying the delay between pulses. We examine two important cases:

\bit
\item \I{On-resonance case:} \zfr{SIfig2}A-B shows the case of $\xt=90^{\circ}$ pulses carried out on resonance. It shows, as expected from basic prethermal dynamics, that the decay time lengthens with increasing pulsing frequency. This can be understood from an increase in effective (time-averaged) Rabi frequency $\Omega$ in this case. The decays themselves exhibit stretched exponential forms with a stretching factor close to 1/2, matching previous experiments \cite{beatrez2021floquet}.

\item \I{Off-resonance case:} \zfr{SIfig2}C-D shows the case of $\xt=5^{\circ}$ pulses applied 5 kHz off-resonance with varying pulsing frequencies. Data shows an increase in lifetime with increase in pulsing frequencies, but the lifetime extension is significantly greater compared to the on-resonance case. The two sharp dips in \zfr{SIfig2}D again correspond to satisfying an effective \(\pi\) condition, where a stable shell-like spin texture forms.
\eit

\subsection{Extended Sensing Data}
In addition to the long-time quantum sensing data shown in \zfr{mfig4} of the main text, here we show several representative time-domain traces which were used to create \zfr{mfig4}(C)-(E). The extended data can be seen in \zfr{SIfig3}. Data here is measured up to $\app$180s. Individual panels display signal amplitude of the $^{13}$C precession signal under simultaneous application of an AC field as a function time. Here, $B_{AC}{=}82\mu$T and $f_{AC}{=}$50Hz as in the main text. The AC field's oscillations are imprinted onto the $^{13}$C precession signal ($\sim 7000$ oscillations per time trace) similar \zfr{mfig4}(B) in the main text, but displayed on the full timescale. Each of the four panels in \zfr{SIfig3} shows the time domain profile for a different frequency offset $\Delta\omega$ and nominal flip angle $\xt$. Surprisingly, when $\Delta\omega{=}$5kHz and $\xt{=}90^{\circ}$, the signal decays extremely quickly as seen in \zfr{SIfig3}(B). However, for the same offset $\Delta\omega{=}$5kHz but shorter nominal flip angle of $\xt{=}15^{\circ}$ as in \zfr{SIfig3}(D), we obtain the best lifetime and are able to detect the external field for $\app$10 mins (as was shown in \zfr{mfig4}(B)(ii)). 

\section{Validation of the Laplace Inversion Fitting}
\zsl{laplace}

As a starting point we assume that the experimental signal $s(t)$ is well-represented by a sum of exponentials with different weights and decay times: 
\begin{equation}
    s(t)=\sum_{j}w_{j}e^{t/T_{2}^{j}}+r(t),
\end{equation}
where $r(t)$ is the residual error of the model. The signal is then suitable for a LI analysis yielding the various weights $w_{j}$ for relaxation times $T_{2}^{j}\in {[}0,{T_{2}^{\rm max}]}$ in a suitable interval. In practice the LI is performed numerically by performing a regularised $\ell^{2}$ optimization on the following objective~\cite{song2002t1, venkataramanan2002solving}
\begin{equation}
    \min_{W}\vert\vert S-KW\vert\vert_{2}^{2} + \alpha\vert\vert W \vert\vert_{2}^{2},
\end{equation}
where $S$ and $W$ are discretized vectors of the signal and weights, respectively. $K_{ij}=e^{-t_{i}/T_{2}^{j}}$ is a matrix representation of the exponential kernel function, and $\alpha$ is a Tikhonov regularization parameter. 
\newline\indent Typically, numerical LI is limited by experimental noise and low sampling rates. However, in our case, these issues are addressed by our high initial polarization levels and high sampling rates of 1GS/s providing a large SNR. 
%Additionally, the time traces $s(t)$ used for the LI optimization contain \mbox{\({>}8 \) M} data points per curve, sampled every $\app$50$\mu$s. 
These factors are unique to our experimental apparatus, enabling us to utilize the full power of the LI algorithm. The resulting fits are robust against noise, and are qualitatively insensitive to changes in the length of measurement. Additionally, the regularization parameter $\alpha$ displays a wide window of acceptable $\alpha$ values that capture the same qualitative information. 

\subsection{Robustness Against Noise}
In \zfr{SIfig4}, we demonstrate the robustness of the LI algorithm against noise using the same data as \zfr{mfig2}A from the main text. \zfr{SIfig4}(A)-(C) shows an on-resonance decay (similar to \zfr{mfig1}(D) from the main text) with different amounts of noise added to it. \zfr{SIfig4}(A) shows a single time domain trace of the raw data, while \zfr{SIfig4}(B)-(C) show single time domain traces of the raw data plus a number randomly sampled from a Gaussian distribution of amplitude 0.01 (\zfr{SIfig4}(B)) or 0.03 (\zfr{SIfig4}(C)) added to each point. \zfr{SIfig4}(A)-(C) is meant to give some intuition to the reader as to how much noise was added to each data set. Below each of these decay curves, \zfr{SIfig4}(D)-(F) shows the corresponding LI map similar to \zfr{mfig2}(A) in the main text. As can be seen in each of the three color plots, the LI map looks extremely similar in each one, therefore demonstrating that the LI algorithm is robust against large amounts of noise added to the data.

\subsection{Effect of Changing Fitting Parameter $\alpha$}
The LI algorithm obtains a fit to the data by minimizing a matrix equation ~\cite{song2002t1, venkataramanan2002solving}. In order to prevent over-fitting such as fitting to noise in the data, a regularization parameter $\alpha$ is used to control the desired smoothness of the fit~\cite{song2002t1, venkataramanan2002solving}. When $\alpha$ is too small, the inversion is unstable in the presence of noise, and when $\alpha$ is too large, the fit will miss some of the finer features in the data. 
In \zfr{SIfig5}, we show the same LI map as the one in \zfr{mfig2}(A) of the main text using a wide range of different $\alpha$ values for the LI algorithm. As can be seen in \zfr{SIfig5}(A), a small $\alpha$ value of 0.001 has a very high resolution but is unstable in the presence of noise and may misinterpret some of the noise in the data as real features. In \zfr{SIfig5}(B), an intermediate $\alpha$ value of 1 prevents over-fitting but still has high enough resolution to capture important features in the data. Finally, in \zfr{SIfig5}(C), a very large $\alpha$ value of 1000 has a low resolution and blurs some of the features that should have been captured by the fitting. In \zfr{SIfig5}(D)-(F), we show that across only 2 orders of magnitude in the value of $\alpha$, the overall qualitative features captured by the LI algorithm are relatively the same but there is some small trade-off between high-resolution and possible over-fitting or lower-resolution and more stability. Typically, the $\alpha$ values used for the LI fits presented in the main text were in the range of 0.1-1. In particular, the $\alpha$ value used in \zfr{mfig2}(A) and \zfr{mfig1} were 0.1. 

\subsection{Effect of Truncating Data}
In \zfr{SIfig6}, we demonstrate the effect of truncating the decay curves to shorter times using the same data from \zfr{mfig2}(A) in the main text. \zfr{SIfig6}(A) shows the LI mapping using the full 10 minute data sets, which is what was done in the main text. In \zfr{SIfig6}(B)-(C), the decay curves used for the LI mapping were truncated to 5 minutes and 3 minutes, respectively. As can be seen in the figure, the qualitative features captured by the LI mapping using the truncated data sets is similar but there is significant blurring of the color map. The main reason why this happens is because when the data is truncated, it is difficult for the LI algorithm to fit very long-lived exponentials to such a short curve.

\section{Spin relaxation under periodic driving}
\zsl{theory}
\subsection{Relaxation model}
A qualitative insight into the relaxation behaviour of the system under periodic driving may be gained following the approach by~\cite{Rhim_calc_1978}, but adopted to the case of non-cyclic evolution. For simplicity we assume an idealised Hamiltonian of the form 
\begin{equation}
    \begin{aligned}
        H_{0}(t)=\omega_{z}\sum_{i}I_{iz}+\omega_{x}(t)\sum_{i}I_{ix}+\sum_{i<j} (d_{ij}-\kappa_{ij}(t)) T^{ij}_{20},
    \end{aligned}
\end{equation}
within the rotating-frame. Here, $\omega_{z}$ is the carrier offset frequency (in rad/s) assumed to be uniform across the ensemble. The radio-frequency field is described by $\omega_{x}(t)$. The secular part of the dipolar interaction is described by the spherical tensor operators $T^{ij}_{20}$ quantised along the space-fixed $z$-axis. In Cartesian coordinates the secular part is given by
\begin{equation}
    \begin{aligned}
T^{ij}_{20}=\frac{1}{\sqrt{6}}(2I_{iz}I_{jz}-(I_{ix}I_{jx}+I_{iy}I_{jy})).
    \end{aligned}
\end{equation}
The dipolar coupling constants $d_{ij}$ are modulated randomly in time via $\kappa_{ij}(t)$. The modulations may be caused by lattice vibrations for example. 

For the current case a $\theta_{x}$ pulse of length $\tau_{1}$ is repeated once every period $T=\tau_{1}+\tau_{2}$
\begin{equation}
    \begin{aligned}
        {[}\theta_{x}(\tau_{1})-\tau_{2}{]}-{[}\theta_{x}(\tau_{1})-\tau_{2}{]}-\dots.
    \end{aligned}
\end{equation}
As a consequence the rf-Hamiltonian is periodic in time
\begin{equation}
    \begin{aligned}
        H_{\rm rf}(t+T)=H_{\rm rf}(t).
    \end{aligned}
\end{equation}
The rf-propagator then admits a Floquet form
\begin{equation}
    \begin{aligned}
        U_{\rm rf}(t)=P(t)e^{-i H_{\rm eff}t}.
    \end{aligned}
\end{equation}
The operator $P(t)$ captures the micro-motion of the system
\begin{equation}
    \begin{aligned}
        P(t+T)=P(t), P(0)=\mathbbm{1}.
    \end{aligned}
\end{equation}
The effective Hamiltonian $H_{\rm eff}$ describes the macro-motion taking the system from $n T$ to $(n+1)T$. The rf-propagator belongs to rotation group SO(3). It is then convenient to choose the Euler representation for $P(t)$
\begin{equation}
    \begin{aligned}
       P(t)=R(\Lambda_{t})=R_{z}(\alpha_{t})R_{y}(\beta_{t})R_{z}(\gamma_{t}).
    \end{aligned}
\end{equation}
The Euler angles are required to be periodic $\Lambda_{t+T}=\Lambda_{t}$ in time. For the effective Hamiltonian we may choose the axis-angle representation
\begin{equation}
    \begin{aligned}
H_{\rm eff}=\omega_{\rm eff} \{R_{z}(\phi_{\rm eff})R_{y}(\theta_{\rm eff})I_{z}R^{\dagger}_{y}(\theta_{\rm eff})R^{\dagger}_{z}(\phi_{\rm eff})\},
    \end{aligned}
\end{equation}
where $(\theta_{\rm eff}, \phi_{\rm eff})$ define a new quantization axis of the system. The effective angles are given by
\begin{equation}
\label{eq:eff_axis}
    \begin{aligned}
&\theta_{\rm eff}=\cot^{-1}\{\cos(\gamma/2)\cot(\eta)+\cot(\psi/2)\csc(\eta)\sin(\gamma/2)\},
\\
&\phi_{\rm eff}=-\frac{\gamma}{2},
    \end{aligned}
\end{equation}
with
\begin{equation}
    \begin{aligned}
\psi=\tau_{1}\sqrt{\omega_{x}^{2}+\omega_{z}^{2}},
\quad\gamma=\tau_{2}\omega_{z},
\quad\eta=\tan^{-1}(\omega_{x},\omega_{z}).
    \end{aligned}
\end{equation}

The relaxation dynamics are analysed within the interaction frame generated by $U_{\rm rf}(t)$. The stochastic part takes the form
\begin{equation}
    \begin{aligned}
Q(t)
&=\sum_{i<j}\kappa_{ij}(t) U^{\dagger}_{\rm rf}(t)T^{ij}_{20}U_{\rm rf}(t)
\\
&=\sum_{i<j}\kappa_{ij}(t) e^{+i H_{\rm eff}t}P^{\dagger}(t)T^{ij}_{20}P(t)e^{-i H_{\rm eff}t},
    \end{aligned}
\end{equation}
\begin{equation}
    \begin{aligned}
Q(t)=\sum_{i<j}\sum_{m}\kappa_{ij}(t) e^{+i H_{\rm eff}t}T^{ij}_{2m}e^{-i H_{\rm eff}t}D^{2}_{m0}(\Lambda^{-1}_{t}),
    \end{aligned}
\end{equation}
where $D^{2}_{mn}(\Lambda)$ are Wigner D-matrix elements. We express the spherical tensor operators $T^{ij}_{2m}$ in terms of spherical tensor operators $B^{ij}_{2m}$ quantized along $n_{\rm eff}$
\begin{equation}
    \begin{aligned}
T^{ij}_{2m}=\sum_{n}B^{ij}_{2n}q_{nm}.
    \end{aligned}
\end{equation}
We may then express $Q(t)$ as follows
\begin{equation}
    \begin{aligned}
Q(t)=\sum_{i<j}d_{ij}(t) \sum_{n,m}B^{ij}_{2n}q_{nm}D^{2}_{m0}(\Lambda^{-1}_{t})e^{+i n\omega_{\rm eff}t}.
    \end{aligned}
\end{equation}
The Wigner D-matrix elements may be expanded as a Fourier series with angular frequency $\Omega=2\pi/T$
\begin{equation}
    \begin{aligned}
Q(t)
&=\sum_{i<j}\kappa_{ij}(t) \sum_{k,n,m}B^{ij}_{2n}q_{nm}c_{mk}e^{+i k\Omega t}e^{+i n\omega_{\rm eff}t}
\\
&=\sum_{i<j}\kappa_{ij}(t) \sum_{k,n}B^{ij}_{2n}(\sum_{m}q_{nm}c_{mk})e^{+i k\Omega t}e^{+i n\omega_{\rm eff}t},
\\
&=\sum_{i<j}\kappa_{ij}(t) \sum_{k,n}B^{ij}_{2n}b_{nk}e^{+i k\Omega t}e^{+i n\omega_{\rm eff}t}.
    \end{aligned}
\end{equation}
For Gaussian statistics the relaxation superoperator in the interaction frame may be written as follows~\cite{ernst_principles_1990,breuer_theory_2007}
\begin{equation}
    \begin{aligned}
\hat{\Gamma}^{I}=-\frac{1}{2}\int_{-\infty}^{\infty}\overline{\hat{Q}(t+\tau)\hat{Q}(t)}d\tau.
    \end{aligned}
\end{equation}
The overline indicates the stochastic average, the hat indicates a commutation superoperator ($\hat{A}B=AB-BA$). Within the rotating-wave approximation we then find
\begin{equation}
\label{eq:Gsop}
    \begin{aligned}
\hat{\Gamma}^{I}\simeq&-\frac{1}{2}\sum_{i,j}\overline{\kappa^{2}_{ij}(0)}\sum_{n,k}\vert b_{nk}\vert^{2}J(n\omega_{\rm eff}+k\Omega)\hat{B}^{ij}_{2n}(\hat{B}^{ij}_{2n})^{\dagger}
\\
=&-\frac{1}{2}\sum_{i,j}\overline{\kappa^{2}_{ij}(0)}\sum_{n}c_{n}\hat{B}^{ij}_{2n}(\hat{B}^{ij}_{2n})^{\dagger},
    \end{aligned}
\end{equation}
where we assume the modulations between two different pairings of spins to be uncorrelated. The spectral density $J(\omega)$ is given by a Lorentzian characterised by a correlation time $\tau_{\rm c}$
\begin{equation}
    \begin{aligned}
J(\omega)=\frac{\tau_{\rm c}}{1+(\omega\tau_{\rm c})^{2}}.
    \end{aligned}
\end{equation}
If we assume that the dipolar interactions are sufficiently averaged by the pulse sequence we may further approximate
\begin{equation}
    \begin{aligned}
&\hat{L}^{I}\simeq \hat{L}^{I}_{\rm dd}+\hat{\Gamma}^{I}, 
\quad 
\hat{L}^{I}_{\rm dd}=-i \sum_{i<j} d_{ij}\hat{B}^{ij}_{20},
    \end{aligned}
\end{equation}
where the first term represents the average dipolar Hamiltonian quantized along the effective axis. The magnetisation after an initial $\pi/2$-pulse may be decomposed with respect to the effective axis $n_{\rm eff}$
\begin{equation}
    \begin{aligned}
\rho^{I}(0)\propto \sum_{m=-1}^{+1}a_{m}(\theta_{\rm eff},\phi_{\rm eff})\sum_{i}p^{i}(0)B^{i}_{1m}.
    \end{aligned}
\end{equation}
The expansion coefficients are given by
\begin{equation}
    \begin{aligned}
&a_{-1}=\frac{1}{\sqrt{2}}(\cos(\theta)\cos(\phi)- i \sin(\phi))=-a^{*}_{+1},
\\
&a_{0}=\sin(\theta)\cos(\phi).
    \end{aligned}
\end{equation}
In general each spin may have a different initial polarisation, but on the time scale of the relaxation process these differences are equalised by spin diffusion effects. In the interaction frame we thus assume equal polarisation for all magnetisation modes
\begin{equation}
    \begin{aligned}
\rho^{I}(t)\simeq \sum_{m=-1}^{+1}a_{m}(\theta_{\rm eff},\phi_{\rm eff})p_{m}(t) \sum_{i}B^{i}_{1m},
    \end{aligned}
\end{equation}
which will approximately relax with rates given by
\begin{equation}
\label{eq:R1eff_analytic}
    \begin{aligned}
R^{'}_{1}&=\frac{(\sum_{i}B^{i}_{10}\vert\hat{\Gamma}^{I}\vert \sum_{i}B^{i}_{10})}{\sum_{i}(B^{i}_{10}\vert B^{i}_{10})}
\\
&=\frac{\Delta M_{2}}{4}\sum_{n}n^2 \sum_{k}\vert b_{nk}\vert^{2}J(n\omega_{\rm eff}+k\Omega),
    \end{aligned}
\end{equation}
\begin{equation}
\label{eq:R2eff_analytic}
    \begin{aligned}
R^{'}_{2}&=\frac{(\sum_{i}B^{i}_{11}\vert\hat{\Gamma}^{I}\vert \sum_{i}B^{i}_{11})}{\sum_{i}(B^{i}_{11}\vert B^{i}_{11})}
\\
&=\frac{\Delta M_{2}}{8}\sum_{n}(6-n^2) \sum_{k}\vert b_{nk}\vert^{2}J(n\omega_{\rm eff}+k\Omega),
    \end{aligned}
\end{equation}
\begin{equation}
    \begin{aligned}
\Delta M_{2}&=\frac{1}{N}\sum_{i<j}\overline{\kappa^{2}_{ij}(0)},
    \end{aligned}
\end{equation}
where $\Delta M_{2}$ is the change in the second moment of the dipolar spectrum. Within the rotating-frame the signal is approximately given by
\begin{equation}
    \begin{aligned}
s(t)\simeq 2\vert a_{1}(\theta_{\rm eff},\phi_{\rm eff})\vert^{2}e^{-R^{'}_{2}t}+\vert a_{0}(\theta_{\rm eff},\phi_{\rm eff})\vert^{2}e^{-R^{'}_{1}t},
    \end{aligned}
\end{equation}
which displays, within the limitations of the model, a bi-exponential decay.

\subsection{Numerical relaxation rates}

%================================================
\begin{figure}
\centering
\includegraphics[trim={0 0 0 0},clip,width=\columnwidth]{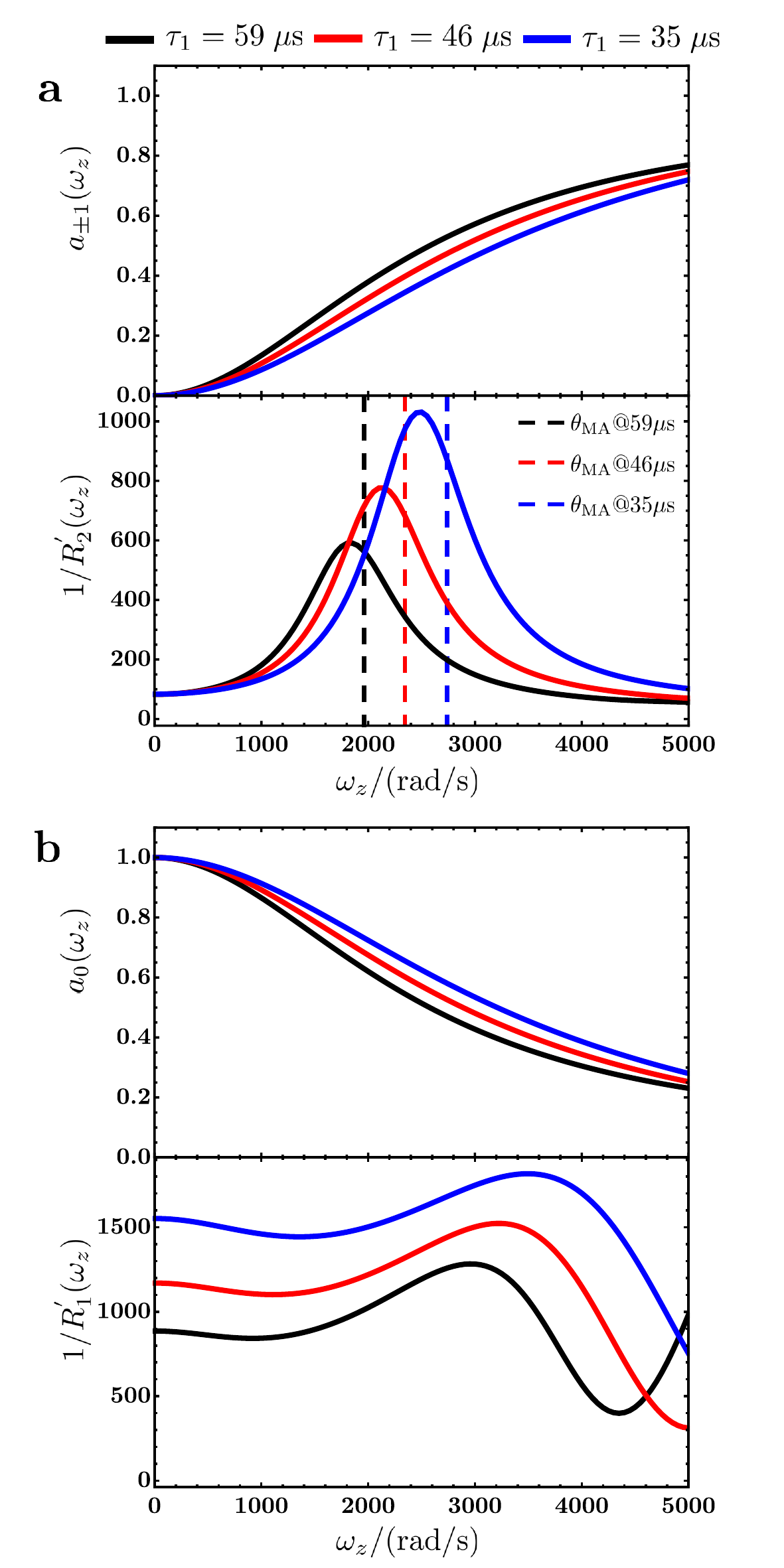}\caption{Relaxation times and relaxation weights for the predicted bi-exponential decay of $\langle I_{x}(t)\rangle$ under periodic spin kicks as a function of the resonance offset ($\omega_{z}$). a) Relaxation rate and weight for transverse magnetisation with respect to the effective quantisation axis. Slight deviations from the exact maxima are expected due to small contributions to $R^{'}_{2}$ from harmonics with $n\neq 2$ in equation \ref{eq:R2eff_analytic}. b) Relaxation rate and weight for longitudinal magnetisation with respect to the effective quantisation axis. Relaxation rates and weights are calculated for varying pulse duration: $\tau_{1}{=}59\;\mu{\rm s}$ (black), $\tau_{1}{=}46\;\mu{\rm s}$ (red), $\tau_{1}{=}35\;\mu{\rm s}$ (blue), but fixed nominal flip angle $\theta_{x}{=}90^{\circ}$. The pulse sequence parameters are chosen in agreement with the experiments: \mbox{$\omega_{x}{=}34$ kHz}, \mbox{$\tau_{c}{=}200\;\mu$s/rad}. Relaxation parameters are given by: \mbox{$\tau_{c}{=}200\;\mu$s/rad}, and \mbox{$\Delta M_{2}{=}2\pi\times 100$ ${\rm(rad/s)}^{2}$}.}
\zfl{R1R2_sim}
\end{figure}
%================================================

The influence of the relaxation components may be explored semi-analytically. To this end the Euler angle trajectory $\Lambda_{t}$ is determined numerically for a single spin-1/2 and a given set of pulse sequence parameters. The numerically constructed trajectory is then utilised for the calculation of the Fourier coefficients $b_{nk}$. \zfr{R1R2_sim} shows the resulting relaxation times $1/R^{'}_{1}$ and $1/R^{'}_{2}$ and the expansion coefficients $a_{m}(\theta_{\rm eff}, \phi_{\rm eff})$ as a function of the resonance offset $\omega_{z}$. \zfr{R1R2_sim}a (top) displays expansion coefficient associated with $R^{'}_{2}$, and indicates an increasing contribution from $R^{'}_{2}$ as the offset increases. Similarly \zfr{R1R2_sim}b (top) shows the expansion coefficient associated with $R^{'}_{1}$, which decreases with increasing offset values. In the offset region $\omega_{z}\in {[}0,3{]}$ kHz the change in $1/R^{'}_{2}$ shows a sharp increase in the relaxation times at $\simeq 1.9$ kHz, $\simeq 2.1$ kHz and $\simeq 2.5$ kHz for pulse length values $\tau_{2}=59\;\mu$s, $\tau_{2}=46\;\mu$s, and $\tau_{2}=35\;\mu$s, respectively. Numerical evaluation of the effective tilt angle $\theta_{\rm eff}$ (equation~\ref{eq:eff_axis}) shows that the maxima occur at $\theta_{\rm eff}\in \{57.35^{\circ}, 57.34^{\circ}, 56.93^{\circ}\}$, which closely align with the magic angle condition. This may also be seen from the analytic expression (equation~\ref{eq:R2eff_analytic}). Since the spectral density is sharply peaked around $J(0)$ the most dominant contributions to the relaxation rate originate from $n=0$ harmonics. These harmonics sample $J(\omega)$ at zero frequency independent of the value of $\omega_{\rm eff}$. If the effective tilt angle is now close to $\theta_{\rm eff}\sim \arccos(1/\sqrt{3})$ the dipolar term $B_{20}$ vanishes and does not contribute to $1/R^{'}_{2}$ leading to a sharp rise in $1/R^{'}_{2}$. 

%At small resonance offsets the relaxation behaviour is mainly determined by $1/R^{'}_{1}$, which is noticeably larger $1/R^{'}_{2}$ within that region. This feature is not visible in the LI data. We attribute this discrepancy to the limitations of the relaxation model, which does not include any mechanisms that effectively relax secular terms. As a result the $1/R^{'}_{1}$ relaxation time is most likely being overestimated by the model. Instead, we expect $1/R^{'}_{1}(\omega_{z})$ to set a fixed relaxation baseline around $1/R^{'}_{2}(0)$. This would explain the significant weight observed in the LI data around $T^{'}_{2}\sim100$ s for $\omega_{z}\lesssim 1$ kHz and the gradual increase for $\omega_{z}\gtrsim 1$ kHz as contributions from $1/R^{'}_{2}$ become more significant.

\subsection{Small angle off-resonant lifetime elongation}

%================================================
\begin{figure}
\centering
\includegraphics[trim={0 0 0 0},clip,width=\columnwidth]{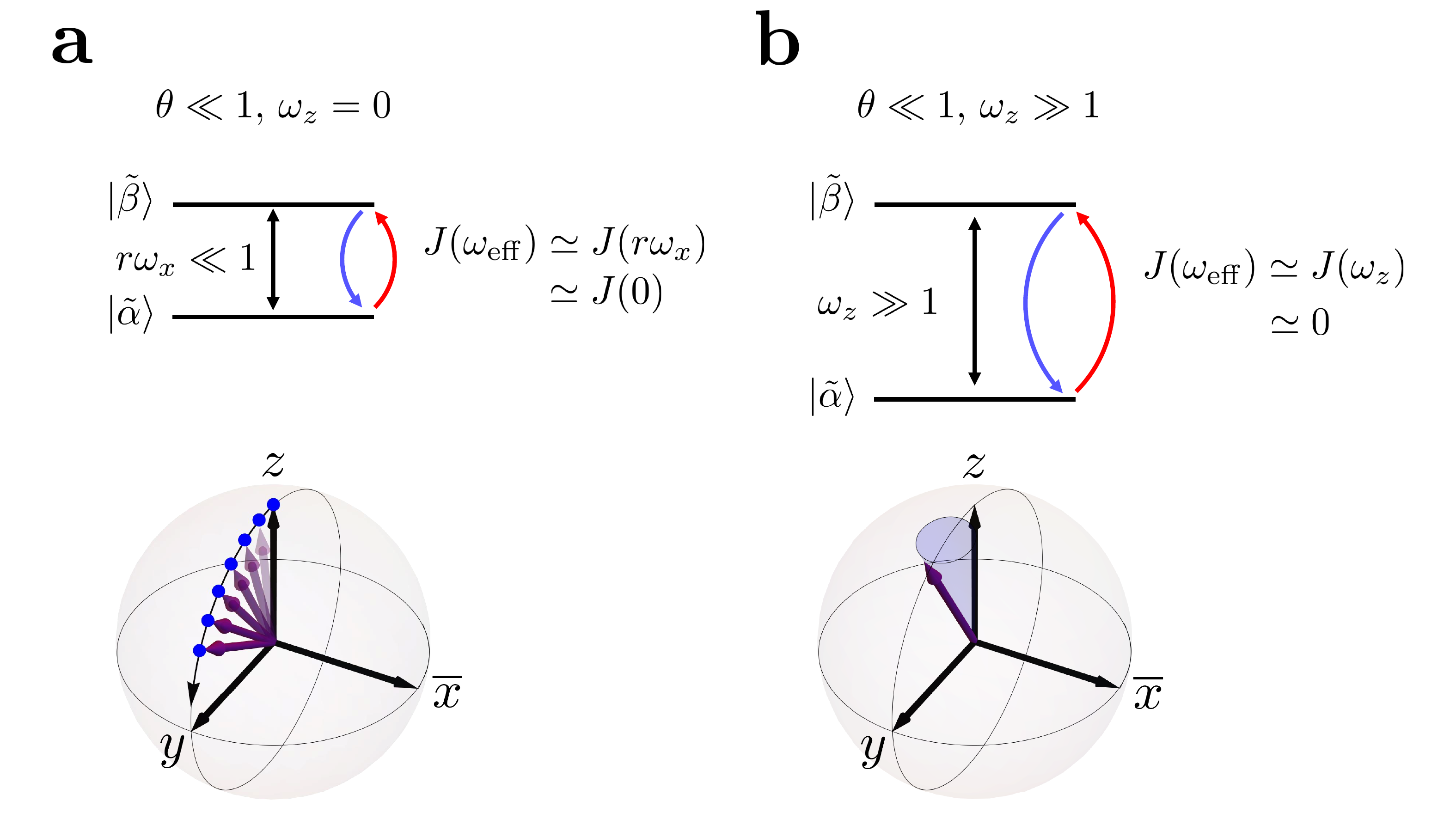}\caption{Schematic picture of anomalous lifetime elongations for small flip angle pulses at distinct resonance conditions for a model two-level system (TLS). Within the interaction frame of the micro-motion operator the energy levels $\tilde{\alpha}$ and $\tilde{\beta}$ of the TLS are determined by the Floquet Hamiltonian. a) For small flip angles applied on resonance the energy gap is given by the time-averaged Rabi frequency $r\omega_{x}{=}\frac{\tau_{1}}{T}\omega_{x}$, which, for finite pulse power, tends towards zero. The transition rate between the Floquet eigenstates becomes proportional to the zero-frequency density $J(0)$, which follows from the fact that even small on-resonance pulses tip the magnetisation through the $xy$-plane. b) For small flip angles applied far off-resonance the energy gap dominated by the detuning $\omega_{z}$. The transition rate between the Floquet eigenstates becomes proportional to density $J(\omega_{z})$ evaluated at the detuning frequency leading to long lifetimes. In contrast to the on-resonance case the spins do not cross the $xy$-plane but undergo conical motion close to the poles.}
\zfl{R1R2_phys}
\end{figure}
%================================================

At small resonance offsets and small flip angles the relaxation behaviour of the longtime component is mainly dominated by $1/R^{'}_{1}$, which tends towards the natural longitudinal relaxation time $1/R_{1}$. A simplified picture of the anomalous lifetime elongations may be given as follows. Within the interaction frame of the micro-motion operator $P(t)$ we have
\begin{equation}
    \begin{aligned}
\tilde{H}(t)=H_{\rm eff}+V(t),
    \end{aligned}
\end{equation}
where the time-independent Floquet Hamiltonian $H_{\rm eff}{=}\omega_{\rm eff} I_{z}$ determines the energy gap between of the system, $V(t)$ represents the random perturbation superimposed with a periodic modulation due to $P(t+T){=}P(t)$. Consider for simplicity the two-level system shown in \zfr{R1R2_phys}. The transition rate between the eigenstates $\tilde{\alpha}$ and $\tilde{\beta}$ depends on the spectral density $J(\omega_{\rm eff}+k 2\pi/T)$ evaluated at the energy gap and the shifted harmonics due to the superimposed periodic modulation. The width of the Lorentzian spectral density however is $w{\sim} 2/\tau_{\rm c}$, which is only $w{\sim}1.6$ kHz for the current case. The experimental sampling rate on the other hand is at least $1/T{>}10$ kHz. The spectral density may thus be considered to be {\em sharply} peaked around $\omega{=}0$ and, as a first approximation, ignore the higher sampling harmonics $k{>}0$. As shown in \zfr{R1R2_phys}a, for the on-resonant case the energy gap is given by the time-averaged Rabi frequency $\omega_{\rm eff}{\simeq} r \omega_{x}$ with $r{=}\tau_{1}/T$. For small flip angles $J(r \omega_{x}){\simeq}J(0)$ leading to efficient relaxation. Intuitively this follows from the fact that even small resonant kicks eventually tip the $z$-magnetisation into the $xy$-plane. Consider now \zfr{R1R2_phys}b, for off-resonant irradiation with small flip angles the energy gap is dominated by $\omega_{z}$. If $\omega_{z}/(2\pi){\gtrapprox}800$ Hz the spectral density essentially vanishes leading to elongated relaxation times. In this case the spins do not cross the $xy$-plane and undergo conical motion close to the poles.

\end{document}